\documentclass[
preprint,
superscriptaddress, prd,tightenlines,showpacs,nofootinbib, eqsecnum,amsfonts,amsmath,amssymb]{revtex4}
\usepackage{bm}
\usepackage{graphicx}
\usepackage{hyperref}
\usepackage{color}

\newcommand{\ud}{\mathrm{d}}
\newcommand{\ab}{^{\alpha\beta}}
\newcommand{\lab}{_{\alpha\beta}}

\newcommand{\R}{{\text{R}}}
\newcommand{\ubar}{{\bar u}}

\newcommand{\uuh}{{\bar{u}^\alpha \bar{u}^\beta h^{\text{R}}_{\alpha\beta}}}

\newcommand{\chsq}{{\chi^2}}

\newcommand{\uT}{u^T}
\newcommand{\SF}{{\text{SF}}}
\newcommand{\hR}{{h^\text{R}_{\alpha\beta}}}

\allowdisplaybreaks
\newcommand{\ua}{^{\alpha}}
\newcommand{\ub}{^{\beta}}
\newcommand{\la}{_{\alpha}}

\newcommand{\beq}{\begin{equation}}
\newcommand{\eeq}{\end{equation}}

\begin{document}

\title{High-Order Post-Newtonian Fit of the Gravitational Self-Force for Circular Orbits in the Schwarzschild Geometry}
\author{Luc Blanchet}\email{blanchet@iap.fr}
\affiliation{$\mathcal{G}\mathbb{R}\varepsilon{\mathbb{C}}\mathcal{O}$, Institut d'Astrophysique de Paris --- UMR 7095 du CNRS, \\ Universit\'e Pierre \& Marie Curie, 98\textsuperscript{bis} boulevard Arago, 75014 Paris, France}
\author{Steven Detweiler}\email{det@phys.ufl.edu}
\affiliation{Institute for Fundamental Theory, Department of Physics, University of Florida, Gainesville, FL 32611-8440, USA}
\author{Alexandre Le Tiec}\email{letiec@iap.fr}
\affiliation{$\mathcal{G}\mathbb{R}\varepsilon{\mathbb{C}}\mathcal{O}$, Institut d'Astrophysique de Paris --- UMR 7095 du CNRS, \\ Universit\'e Pierre \& Marie Curie, 98\textsuperscript{bis} boulevard Arago, 75014 Paris, France}
\author{Bernard F. Whiting}\email{bernard@phys.ufl.edu}
\affiliation{Institute for Fundamental Theory, Department of Physics, University of Florida, Gainesville, FL 32611-8440, USA}

\date{\today}

\begin{abstract}
We continue a previous work on the comparison between the post-Newtonian (PN) approximation and the gravitational self-force (SF) analysis of circular orbits in a Schwarzschild background. We show that the numerical SF data contain physical information corresponding to extremely high PN approximations. We find that knowing analytically determined appropriate PN parameters helps tremendously in allowing the numerical data to be used to obtain higher order PN coefficients. Using standard PN theory we compute analytically the leading 4PN and the next-to-leading 5PN logarithmic terms in the conservative part of the dynamics of a compact binary system. The numerical perturbative SF results support well the analytic PN calculations through first order in the mass ratio, and are used to accurately measure the 4PN and 5PN non-logarithmic coefficients in a particular gauge invariant observable. Furthermore we are able to give estimates of higher order contributions up to the 7PN level. We also confirm with high precision the value of the 3PN coefficient. This interplay between PN and SF efforts is important for the synthesis of template waveforms of extreme mass ratio inspirals to be analysed by the space-based gravitational wave instrument LISA.  Our work will also have an impact on efforts that combine numerical results in a quantitative analytical framework so as to generate complete inspiral waveforms for the ground-based detection of gravitational waves by instruments such as LIGO and Virgo.
\end{abstract}


\maketitle

\section{Motivation and summary}\label{secI}

This paper is the follow up of previous work \cite{Bl.al.10} (hereafter Paper I) where we demonstrated a very good agreement between the analytical post-Newtonian (PN) approximation and the numerical gravitational self-force (SF) for circular orbits in the perturbed Schwarzschild geometry. The first step had been taken by Detweiler \cite{De.08} who showed agreement at 2PN order using the existing PN metric \cite{Bl.al.98}.\footnote{As usual the $n$PN order refers to terms equivalent to $(v/c)^{2n}$ beyond Newtonian theory, where $v$ is a typical internal velocity of the material system and $c$ is the speed of light.} Motivated by this result we pushed the PN calculation in Paper I up to the 3PN level. This is particularly interesting because the 3PN approximation necessitates an extensive use of dimensional regularization to treat the divergent self-field of point particles. The successful comparison reported in Paper I confirmed the soundness of both the traditional PN expansion (see e.g. \cite{Bl.06}) and the perturbative SF analysis \cite{Mi.al.97,QuWa.97,DeWh.03,GrWa.08,Po.04} in describing the dynamics of compact binary systems --- notably, regarding subtleties associated with the self-field regularizations in use in both approaches. This comparison dealt with the conservative part of the dynamics, but previous comparisons between the PN and the SF had checked dissipative effects \cite{Po.93,Cu.al2.93,Po2.93,TaNa.94,Po.95,Ta.al.96,Ta.al2.96,PoSa.95}. 

In Paper I we also showed that the quality of the numerical SF data is such that substantial physical information remains far beyond 3PN order, i.e. is contained within the numerically derived residuals obtained after subtracting the known 3PN terms from the data (see Fig.~3 of Paper I). In the present paper we explore further the higher-order PN nature of the numerical data. We point out that knowing analytically determined appropriate PN parameters helps tremendously in allowing our numerical data to be used to obtain higher order PN terms. In particular, we show that prior analytic information from PN theory regarding the presence of \textit{logarithmic} terms in the PN expansion is crucial for efficiently extracting from the SF data the numerical values of higher order PN coefficients. 

The occurence of logarithmic terms in the PN expansion has been investigated in many previous works \cite{And,Ke1.80,Ke2.80,An.al.82,FuSc.83,Fu.83,BlDa.86,BlDa.88}. Notably Anderson \textit{et al.} \cite{An.al.82} found that the dominant logarithm arises at the 4PN order, and Blanchet \& Damour \cite{BlDa.88} (see also \cite{Bl.93}) showed that this logarithm is associated with gravitational wave tails modifying the usual 2.5PN radiation-reaction damping at the 1.5PN relative order. Furthermore the general structure of the PN expansion is known \cite{BlDa.86}: it is of the type $\sum (v/c)^k[\ln(v/c)]^q$, where $k$ and $q$ are positive integers, involving only powers of logarithms; more exotic terms such as $[\ln(\ln(v/c))]^q$ cannot arise. In the present paper we shall determine the leading 4PN logarithm and the next-to-leading 5PN logarithm in the conservative part of the dynamics of a compact binary system.

Consider two compact objects with masses $m_1$ and $m_2$ (without spins) moving on exactly circular orbits. The dissipative effects associated with gravitational wave emission are neglected, which is formalized by assuming the existence of a helical Killing vector field $K\ua(x)$, being null on the light cylinder associated with the circular motion, time-like inside the light cylinder (for instance at the particle's location) and space-like outside (including a neighborhood of spatial infinity). Then we consider a particular gauge invariant observable quantity \cite{De.08} defined as the constant of proportionality between the four-velocity of one of the masses, say $m_1$, and the helical Killing vector evaluated at the location of that particle, i.e. $K_1\ua\equiv K\ua(y_1)$, 
\begin{equation}\label{uTdef}
   u_1\ua = u_1^T \,K_1\ua\,.
\end{equation}
The quantity $u_1^T$ represents the redshift of light rays emitted from the particle and received on the helical symmetry axis perpendicular to the orbital plane \cite{De.08}; we shall sometimes refer to it as the redshift observable. Adopting a coordinate system in which the helical Killing vector field reads $K\ua\partial\la = \partial_t + \Omega\,\partial_\varphi$, where $\Omega$ denotes the orbital frequency of the circular motion, we find that the redshift observable reduces to the $t$ component $u_1^t\equiv\ud t/\ud\tau_1$ of the particle's four-velocity, namely
\begin{equation}\label{uT}
    u_1^T = u_1^t = \biggl( - g\lab(y_1) \frac{v_1\ua v_1\ub}{c^2} \biggr)^{-1/2} \,.
\end{equation}
Here $v_1^\alpha\equiv\ud y_1^\alpha/\ud t=(c,v_1^i)$ is the ordinary coordinate velocity used in PN calculations, and $g\lab(y_1)$ denotes the metric being evaluated at the particle's location by means of an appropriate self-field regularization, i.e. mode-sum regularization in the SF approach, and dimensional regularization in the PN context. 

The point is that $u_1^T$ can be computed as a function of the orbital frequency $\Omega$ in both the PN approach for any mass ratio, and in the perturbative SF framework when the mass $m_1$ is much smaller than $m_2$. Summarizing the analytical 3PN result of Paper I and present computation of the 4PN and 5PN logarithmic terms in Secs.~\ref{secII}--\ref{secV}, we obtain the SF contribution to the redshift observable \eqref{uT} as\footnote{Inspired by our earlier work \cite{Bl.al.10}, the easy calculation of the 4PN logarithm has already been given in \cite{Da.10}.}
\begin{align}\label{utSFintro}
u^T_{\mathrm{SF}} =& - y - 2 y^2 - 5 y^3 + \left(
- \frac{121}{3} + \frac{41}{32} \pi^2 \right) y^4 \nonumber\\&+ \left(
\alpha_4 - \frac{64}{5}\ln y\right) y^5 + \left(
\alpha_5 + \frac{956}{105}\ln y\right) y^6 + o(y^6)\,,
\end{align}
where $y=(G m_2 \Omega/c^3)^{2/3}$ is a PN parameter associated with the lighter mass $m_1$, and $\alpha_4$ and $\alpha_5$ denote some purely numerical coefficients left out in the PN calculation. However, having obtained theoretical predictions for the 4PN and 5PN logarithmic terms, we are able to perform an efficient fit to the numerical SF data and to accurately measure the other non-logarithmic 4PN and 5PN coefficients. We find $\alpha_4=-114.34747(5)$ and $\alpha_5=-245.53(1)$ where the uncertainty in the last digit is in parenthesis. Furthermore we can also measure the 6PN coefficients $\alpha_6$ and $\beta_6$ (such that $\alpha_6+\beta_6\ln y$ is the factor of $y^7$), and give an estimate of the total contribution of the 7PN coefficient (including both logarithmic and non-logarithmic terms); see Table \ref{bestfit} and Fig. \ref{bestfig} in Sec.~\ref{secVID}. The 3PN coefficient $\alpha_3 = - \frac{121}{3} + \frac{41}{32} \pi^2$ is also found to be in agreement with the SF data with high precision.

The non-logarithmic coefficients $\alpha_4$, $\alpha_5$, $\cdots$ would be extremely difficult to obtain with standard PN methods. Their computation would require in particular having a consistent self-field regularization scheme; for instance it is not guaranteed that dimensional regularization which has been so successful at 3PN order could be applied with equal success at much higher orders. Nevertheless these coefficients are obtained here for the first time with reasonable precision up to the impressive 7PN order. This emphasizes the powerfulness of the perturbative SF approach and its ability to describe the strong field regime of compact binary systems, which is inaccessible to the PN method. Of course, the limitation of the SF approach is the small mass-ratio limit; in this respect it is taken over by the PN method.

The analytical and numerical results obtained in this paper up to 7PN order could be used for the synthesis and calibration of template waveforms of extreme mass ratio inspirals to be observed by the space-based gravitational wave detector LISA. They are also relevant to analyses that combine numerical computations in a quantitative analytical framework for the generation of inspiral waveforms for the ground-based LIGO and Virgo detectors.

The remainder of this paper is organized as follows: In Sec.~\ref{secII} we perform a detailed analysis of the occurence of logarithmic terms in the near-zone expansion of an isolated source. This general discussion is followed in Sec.~\ref{secIII} by the explicit computation of the leading order 4PN and next-to-leading order 5PN logarithmic terms in the near-zone metric of an arbitrary post-Newtonian source, and then of a compact binary system. We proceed in Sec.~\ref{secIV} with the computation of these terms in the acceleration of the compact binary, as well as in the binary's conserved energy, and consider the restriction to circular orbits. This allows us to derive intermediate results necessary for the computation of the 4PN and 5PN logarithmic terms in the redshift observable \eqref{uT} for circular orbits; this is detailed in Sec.~\ref{secV}. Finally, Sec.~\ref{secVI} is devoted to a high-order PN fit of our numerical data for the SF effect on the redshift variable. The Appendix provides general formulas for the computation of logarithmic terms in PN theory.

\section{General structure of logarithmic terms}\label{secII}

\subsection{Near-zone expansion of the exterior metric}
\label{secIIA}

In this Section we study in a general way the PN orders at which logarithmic terms occur in the near-zone expansion of the metric of an isolated source. Our main tool will be the multipolar-post-Minkowskian (MPM) analysis of the vacuum field outside the compact support of the source \cite{BlDa.86,BlDa.88,Bl.93,Bl.98,PoBl.02}. The starting point is the general solution of the linearized vacuum Einstein field equations in harmonic coordinates, which takes the form of a multipolar expansion parametrized by mass-type $M_L$ and current-type $S_L$ multipole moments \cite{Th.80}\footnote{Our notation is as follows: $L = i_1 \cdots i_\ell$ denotes a multi-index composed of $\ell$ multipolar spatial indices $i_1, \cdots, i_\ell$ (ranging from 1 to 3); $\partial_L = \partial_{i_1} \cdots \partial_{i_\ell}$ is the product of $\ell$ partial derivatives $\partial_i \equiv \partial / \partial x^i$; $x_L = x_{i_1} \cdots x_{i_\ell}$ is the product of $\ell$ spatial positions $x_i$; similarly $n_L = n_{i_1} \cdots n_{i_\ell}$ is the product of $\ell$ unit vectors $n_i=x_i/r$; the symmetric-trace-free (STF) projection is indicated with a hat, i.e. $\hat{x}_L \equiv \text{STF}[x_L]$, $\hat{n}_L \equiv \text{STF}[n_L]$, $\hat{\partial}_L \equiv \text{STF}[\partial_L]$, or sometimes using brackets surrounding the indices, i.e. $x_{\langle L \rangle} \equiv \hat{x}_L$. In the case of summed-up (dummy) multi-indices $L$, we do not write the $\ell$ summations from 1 to 3 over their indices. The totally antisymmetric Levi-Civita symbol is denoted $\varepsilon_{ijk}$; symmetrization over indices is denoted $(ij)=\frac{1}{2}(ij+ji)$; time-derivatives of the moments are indicated by superscripts $(n)$.} 
\begin{subequations}\label{h1}
\begin{align}
    h^{00}_{1} &= -\frac{4}{c^2}\sum_{\ell\geqslant 0}
      \frac{(-)^\ell}{\ell !} \partial_L \left[ \frac{1}{r} M_L(u)
      \right]\,, \label{h100}\\ h^{0i}_{1} &= 
  \frac{4}{c^3}\sum_{\ell\geqslant 1} \frac{(-)^\ell}{\ell !} \left\{
      \partial_{L-1} \left[ \frac{1}{r} M_{iL-1}^{(1)}(u)
      \right] + \frac{\ell}{\ell+1} \varepsilon_{iab} \partial_{aL-1}
      \left[ \frac{1}{r} S_{bL-1}(u) \right]\right\}\,, \\ 
h^{ij}_{1} &= -\frac{4}{c^4}\sum_{\ell\geqslant 2}
      \frac{(-)^\ell}{\ell !} \left\{ \partial_{L-2} \left[ \frac{1}{r}
      M_{ijL-2}^{(2)}(u) \right] + \frac{2\ell}{\ell+1}
      \partial_{aL-2} \left[ \frac{1}{r} \varepsilon_{ab(i}
      S_{j)bL-2}^{(1)}(u) \right]\right\}\,.
  \end{align}
\end{subequations}
The multipole moments $M_L$ and $S_L$ are symmetric and trace-free (STF) with respect to all their indices and depend on the retarded time $u\equiv t-r/c$ in harmonic coordinates. They describe a general isolated source and are unconstrained except that the mass monopole $M$ and current dipole $S_i$ are constant, and the mass dipole $M_i$ is varying linearly with time.

Starting from $h_{1}$ we define a full non-linear MPM series for the ``gothic'' metric deviation $h\ab\equiv \sqrt{-g} \, g\ab-\eta\ab$ (where $g\ab$ and $g$ denote the inverse and determinant of the usual covariant metric respectively, and where $\eta\ab$ is the Minkowski metric) as
\begin{equation}\label{MPM}
h\ab = \sum_{n=1}^{+\infty}G^n h_{n}\ab\,,
\end{equation}
where the Newton constant $G$ serves at labelling the successive post-Minkowskian orders. Plugging this series into the (vacuum) Einstein field equations in harmonic coordinates we find at each order $\partial_\mu h_{n}^{\alpha\mu} = 0$, together with
\begin{equation}\label{EEn}
\Box h_{n}\ab = N_{n}\ab\,,
\end{equation}
where $\Box=\eta^{\mu\nu}\partial_{\mu}\partial_{\nu}$ is the flat d'Alembertian operator, and where $N_{n}$ denotes the $n$-th non-linear gravitational source term depending on previous iterations $h_{1}$, $\cdots$, $h_{n-1}$. An explicit ``algorithm'' has been proposed in \cite{BlDa.86} for solving \eqref{EEn} and the condition of harmonic coordinates at any post-Minkowskian order $n$.

We are interested in the expansion of the solution of \eqref{EEn} in the near-zone (NZ), i.e. formally when $r\rightarrow 0$ (but still outside the compact supported source). The general structure of that expansion is known \cite{Bl.93}. For the source term we have (the NZ expansion being indicated with an overbar)
\begin{equation}\label{Nexp}
\overline{N}_{n}\ab = \sum_{E_n}\frac{1}{c^{3n+\sum_{i=1}^n\underline{\ell}_i+2}}\sum_{\ell, p, q\atop q\leqslant n-2}F_{Lpq}\ab(t)\,\hat{n}_L\left(\frac{r}{c}\right)^p\left[\ln\left(\frac{r}{\lambda}\right)\right]^q\,.
\end{equation}
We see that besides the normal powers of $r$ we have also powers of logarithms of $r$; $p$ is an integer ($p\in\mathbb{Z}$) bounded from below by some $p_0$ depending on $E_n$, and $q$ is a positive integer ($q\in\mathbb{N}$). We pose $\lambda=2\pi c/\Omega$, with $\Omega$ a typical frequency scale in the source to be identified later with the orbital frequency of the binary's circular orbit. We denote by $E_n=\{M_{L_1},M_{L_2},\cdots,\varepsilon_{ai_{\ell_n+1}i_{\ell_n}}S_{aL_n-1}\}$ a set of $n$ multipole moments, with the current moments endowed with their natural Levi-Civita symbol. We pose $\underline{\ell}_i=\ell_i$ for mass moments and $\underline{\ell}_i=\ell_i+1$ for current moments, so that $\sum_{i=1}^n\underline{\ell}_i$ is the total number of indices carried by the moments of the set $E_n$. On the other hand $\ell$ is the number of indices carried by the STF multipolar factor $\hat{n}_L$. The multipole functions in \eqref{Nexp} admit the general structure \cite{Bl.93}
\begin{equation}\label{FLpq}
F_{Lpq}\ab(t)=\int\ud u_1\cdots\int\ud u_n\,\mathcal{K}_{L\underline{L}_1\cdots\underline{L}_n}\ab(t,u_1,\cdots,u_n) \, M_{L_1}^{(a_1)}(u_1)\cdots\varepsilon_{ai_{\ell_n+1}i_{\ell_n}}S_{aL_n-1}^{(a_n)}(u_n)\,,
\end{equation}
where the kernel $\mathcal{K}$ has an index structure made only of Kronecker symbols and is only a function of time variables: the current time $t$, the $n$ integration arguments $u_i$ (satisfying $u_i\leqslant t$) and the period $P=\lambda/c$ of the source. Then with this convention we see that the powers of $1/c$ in \eqref{Nexp} are set by dimensionality. A useful lemma \cite{Bl.93} is the fact that the multipolar order $\ell$ is necessarily constrained by the following two inequalities:
\begin{equation}\label{lemma}
-\sum_{i=1}^n\underline{\ell}_i+4-s\,\leqslant\,\ell\,\leqslant\,\sum_{i=1}^n\underline{\ell}_i+s\,.
\end{equation}
Here $s$ is the number of spatial indices among $\alpha$ and $\beta$, i.e. the ``spin'' given by $s=0,1,2$ according to $\alpha\beta=00,0i,ij$. 

The lemma \eqref{lemma} will serve at controlling the PN order of ``branches'' of logarithmic terms arising in the MPM iteration of the external field. Already we know \cite{BlDa.86} that the powers of the logarithms are limited to $q\leqslant n-2$ in the source term $\overline{N}_{n}$. After integration of the source term $\overline{N}_{n}$ we shall find the corresponding solution $\overline{h}_{n}$ which will admit the same type of NZ expansion as its source. However the maximal power of the logarithms in the solution will be increased by one unit with respect to the source and is thus limited by $n-1$, i.e. $q\leqslant n-1$ in $\overline{h}_{n}$. For instance this means that logarithms squared cannot arise before the cubic non-linear order $n=3$.

To control the occurence of logarithms in the near-zone it will be sufficient to integrate the source \eqref{Nexp} by means of the integral of the ``instantaneous'' potentials defined by formal PN iteration of the inverse Laplace operator $\Delta^{-1}$, say $\Box^{-1}=\Delta^{-1}+c^{-2}\partial_t^2\Delta^{-2}+\cdots$. This is because any homogeneous solution to be added to that particular solution will have the structure of a free multipolar wave (retarded or advanced) whose near-zone expansion cannot contain any logarithms. However, when acting on a multipolar expanded source term, valid only in the exterior of the matter source and becoming singular in the formal limit $r\rightarrow 0$, we must multiply the source term by a regulator $(r/\lambda)^B$, where $B$ is a complex number and $\lambda=cP$ is the length scale associated with the orbital motion. After applying the instantaneous propagator we take the finite part (FP) of the Laurent expansion when $B\rightarrow 0$. Thus the solution reads as
\begin{equation}\label{hexp}
\overline{h}_{n}\ab = \mathop{\mathrm{FP}}_{B=0}\,\sum_{k=0}^{+\infty}\left(\frac{\partial}{c\partial t}\right)^{2k}\!\!\Delta^{-k-1}\left[\left(\frac{r}{\lambda}\right)^B \overline{N}_{n}\ab\right] + \overline{H}_{n}\ab\,.
\end{equation}
Later, in \eqref{instpot} below, we shall denote by $\mathcal{I}^{-1}$ the particular ``instantaneous'' regularized propagator appearing in \eqref{hexp}. The term $\overline{H}_{n}$ denotes the NZ expansion of an homogeneous solution of the d'Alembert equation. In the general case this solution will be a mixture of purely retarded and advanced multipolar waves, say of the type $\sum\hat{\partial}_L\{R_L(t-r/c)/r\}$ and $\sum\hat{\partial}_L\{A_L(t+r/c)/r\}$, but the point is that the NZ expansion of $\overline{H}_{n}$ when $r\rightarrow 0$ clearly does not contain any logarithms. So in order to control the logarithms we can ignore the homogeneous piece $\overline{H}_{n}$.

As argued in \cite{Bl.93} the use of the latter ``instantaneous'' propagator, say $\mathcal{I}^{-1}$, corresponds to keeping only the conservative part of the dynamics, i.e. neglecting the dissipative part associated with gravitational radiation-reaction. Below we shall implement the restriction to the conservative case by looking at circular orbits with helical Killing symmetry. We expect that a solution admitting this symmetry should be given by \eqref{hexp} where the homogeneous part $\overline{H}_{n}$ is of the symmetric type $\sum\hat{\partial}_L\{[S_L(t-r/c)+S_L(t+r/c)]/r\}$. In this ``symmetric'' situation, where the radiation-reaction is neglected, the solution should depend on the length scale $\lambda$ appearing in the first term of \eqref{hexp}. Indeed this length scale is introduced in the problem by our assumption of having the helical Killing symmetry with Killing vector $K\ua\partial\la = \partial_t + \Omega\,\partial_\varphi$ where $\Omega=2\pi c/\lambda$.

\subsection{Near-zone versus far-zone logarithms}
\label{secIIB}

Inserting the general form of the source term \eqref{Nexp} into \eqref{hexp}, and ignoring from now on the homogeneous term $\overline{H}_{n}$ which does not contain logarithms, we obtain (dropping the space-time indices $\alpha\beta$ for clarity)
\begin{equation}\label{hexp2}
\overline{h}_{n} =\sum_{E_n}\frac{1}{c^{3n+\sum_{i=1}^n\underline{\ell}_i+2}}\sum_{\ell, p, q\atop q\leqslant n-2} \sum_{k=0}^{+\infty}
\frac{F_{Lpq}^{(2k)}(t)}{c^{p+2k}}\mathop{\mathrm{FP}}_{B=0}\Delta^{-k-1}\left\{\left(\frac{r}{\lambda}\right)^B\hat{n}_L \,r^{p}\left[\ln\left(\frac{r}{\lambda}\right)\right]^q\right\}\,.
\end{equation}
We can explicitly integrate the iterated Poisson integral and find
\begin{equation}\label{mathieu}
\Delta^{-k-1}\left\{\left(\frac{r}{\lambda}\right)^B\hat{n}_L \,r^{p}\left[\ln\left(\frac{r}{\lambda}\right)\right]^q\right\} = \left(\frac{\partial}{\partial B}\right)^q\left[\alpha_{\ell, p, k}(B)\left(\frac{r}{\lambda}\right)^B\hat{n}_L \,r^{p+2+2k}\right]\,,
\end{equation}
with $B$-dependent coefficients
\begin{equation}\label{alphaB}
\alpha_{\ell, p, k}(B)=\prod_{i=0}^{k}\frac{1}{(B+p+2+2i-\ell)(B+p+3+2i+\ell)}\,.
\end{equation}

We shall now control the occurence of a pole $\propto 1/B$ in the latter expression which, after taking the finite part in \eqref{hexp2}, will generate a logarithm $\ln r$. Actually, since we have to differentiate $q$ times with respect to $B$, the pole in $\alpha_{\ell, p, k}(B)$ (which is necessarily a simple pole) will yield multiple poles $\propto 1/B^{m}$, and we shall finally end up with powers of logarithms $(\ln r)^{m}$, where here $m\leqslant q+1$ --- hence the increase by one of the powers of logarithms from the source to the solution, as discussed previously.

Inspection of Eq.~\eqref{alphaB} readily shows that there are two types of poles. First we have the poles for which $p+2=\ell-2i$. These will be qualified as ``near-zone poles'', and the structure of the solution for these poles reads 
\begin{equation}\label{NZpole}
\left(\overline{h}_{n}\right)_\text{NZ pole} = \sum_{\ell,j\geqslant 0\atop m\leqslant n-1}\frac{1}{c^{3n+\sum_{i=1}^n\underline{\ell}_i+\ell}}\,G_{Ljm}(t)\,\hat{x}_L\left(\frac{r}{c}\right)^{2j}\left[\ln\left(\frac{r}{\lambda}\right)\right]^m\,,
\end{equation}
where $j=k-i\geqslant 0$ and the functions $G_{Ljm}(t)$ have a structure similar to \eqref{FLpq}. Note that \eqref{NZpole} is perfectly regular when $r\rightarrow 0$ [at least when $\ell+j\geqslant 1$] and will therefore be valid (after matching) inside the matter source. On the other hand the ``far-zone poles'' for which $p+2=-\ell-1-2i$ have the structure
\begin{equation}\label{FZpole}
\left(\overline{h}_{n}\right)_\text{FZ pole} = \sum_{\ell,j\geqslant 0\atop m\leqslant n-1}\frac{1}{c^{3n+\sum_{i=1}^n\underline{\ell}_i-\ell-1}}\,K_{Ljm}(t)\,\hat{\partial}_L\!\left(\frac{1}{r}\right)\left(\frac{r}{c}\right)^{2j}\left[\ln\left(\frac{r}{\lambda}\right)\right]^m\,.
\end{equation}
These poles become singular when $r\rightarrow 0$. We shall argue later that the associated logarithms do not contribute to the PN expansion of quantities we compute in this paper (like the redshift observable or the conserved energy of a compact binary system).

We can now easily control the PN order of these poles. Taking into account all the powers of $1/c$ and the fact that $j\geqslant 0$, we obtain
\begin{equation}\label{NZordre0}
\left(\overline{h}_{n}\right)_\text{NZ pole} = \sum_\ell \mathcal{O}\left(\frac{1}{c^{3n+\sum_{i=1}^n\underline{\ell}_i+\ell}}\right)\,.
\end{equation}
Next, the inequality in the left of the lemma \eqref{lemma} provides a uniform bound of the PN order of each of the terms in \eqref{NZordre0}, leading to
\begin{equation}\label{NZordre}
\left(\overline{h}_{n}\right)_\text{NZ pole}  = \mathcal{O}\left(\frac{1}{c^{3n+4-s}}\right)\,.
\end{equation}
This means that the NZ poles in the $n$-th non-linear metric are produced at least at the $\frac{3n+2}{2}$PN level; note that the power of $1/c$ in the $ij$ components of the perturbation $\overline{h}_{n}$, such that $s=2$, gives immediately the PN order. Similarly we find, using now the inequality on the right of \eqref{lemma}, that the FZ poles are produced at the level
\begin{equation}\label{FZordre}
\left(\overline{h}_{n}\right)_\text{FZ pole} = \mathcal{O}\left(\frac{1}{c^{3n-1-s}}\right)\,,
\end{equation}
corresponding to (at least) the $\frac{3n-3}{2}$PN order. Notice that the far-zone poles come earlier than the near-zone ones in the PN iteration.

We use these general results to control the occurence of (powers of) logarithms in the PN expansion. First be careful that our findings do not mean that all the logarithms at some $n$-th non-linear order will have the PN orders \eqref{NZordre} and \eqref{FZordre}; it states that whenever \textit{new logarithms} appear they are necessarily produced at least at these PN levels. However, once a ``new'' logarithm has been produced in $\overline{h}_{n}$, it will contribute in the source term $\overline{N}_{n+1}$ of the next iteration, and therefore will also appear in the corresponding solution $\overline{h}_{n+1}$ where it needs not to be associated with a pole occuring at that order. In fact we expect that the vast majority of logarithms only come from the iteration of original logarithms seeded by poles. Such ``iterated'' logarithms will escape the rules \eqref{NZordre} and \eqref{FZordre}. 

Given a logarithm at order $n$ coming from a NZ pole and being thus at least of order $\frac{3n+2}{2}$PN, we can check that it will generate iterated logarithms at any subsequent non-linear order $n+p$, with $p\geqslant 1$, and that those will be at least of order $\frac{3n+2p}{2}$PN. We can therefore always bound the PN order of the complete family of iterated NZ logarithms by the order $\frac{3n+2}{2}$PN of the ``seed'' logarithm.\footnote{When $p=1$ we get the same PN order as the seed logarithm because according to \eqref{NZordre} the $ij$ component of the metric perturbation $h_n\ab$ is of order $1/c^{3n+2}$, and hence generates at the next iteration a term of order $1/c^{3n+4}$ in the 00 component of the metric perturbation $h_{n+1}\ab$ (via the non-linear source term $h_n^{ij}\partial_{i}\partial_{j}h_1^{00}$), which is still of $\frac{3n+2}{2}$PN order. We shall use later the trick that by gauging away the $ij$ component of the metric perturbation we can greatly simplify the computation of the subsequent iteration.} The same reasoning applies for the PN orders of the iterated FZ logarithms which are bounded from below by the $\frac{3n-3}{2}$PN order of the seed.

When $n=2$ we find from \eqref{NZordre} that there is a family of NZ logarithms starting at the 4PN order. We know that the 4PN logarithmic term is associated with gravitational wave tails; it has been computed for general matter sources in \cite{BlDa.88}. Conjointly with this 4PN logarithm there will be also logarithms at 5PN and higher orders, all of them at quadratic order $n=2$, and all these quadratic logarithms will have to be iterated at the next cubic order $n=3$, and so on. As we discussed this defines a complete family of NZ logarithms, and this family will be sufficient to control all the NZ logarithms at 4PN and 5PN orders. Indeed, we expect that at cubic order $n=3$ a new family of NZ logarithms will appear, but according to the result \eqref{NZordre} this new family will be of order 5.5PN at least. In particular this reasonning shows that the dominant NZ logarithm \textit{squared} $[\ln(r/\lambda)]^2$ is at least 5.5PN order. Such 5.5PN logarithm would be time odd in a time reversal and belongs to the dissipative radiation-reaction part of the dynamics so we shall ignore it. Similarly the next family coming at the quartic approximation $n=4$ will be at least 7PN --- thus the dominant $[\ln(r/\lambda)]^3$ is expected to appear at least at 7PN order. 

We shall now argue that only the family of NZ logarithms starting at the 4PN order needs to be considered for the present computation, because quite generally the FZ logarithms cannot contribute to the conserved part of the dynamics of a compact binary system.

\subsection{Argument that far-zone logarithms give zero contribution}
\label{secIIC}

The FZ logarithms are generated by seeds whose PN order is controlled by the estimate \eqref{FZordre}. First one can check that due to the particular structure of the quadratic metric $n=2$ there is no FZ pole at the quadratic order \cite{BlDa.88}. The FZ logarithms come only at the cubic order $n=3$ and from the estimate \eqref{FZordre} we see that they arise dominantly at 3PN order, i.e. earlier than the NZ logarithms at 4PN order. The 3PN far-zone logarithms have been investigated in \cite{BlDa.88} and also in previous work \cite{An.al.82}. However we do not need to consider these and other FZ logarithms in the present calculation as the following argument shows.

The NZ and FZ logarithms were investigated using the operator of the ``instantaneous'' potentials defined by [see Eq.~\eqref{hexp}]
\begin{equation}\label{instpot}
\mathcal{I}^{-1}\left[\overline{N}_{n}\right] = \mathop{\mathrm{FP}}_{B=0}\,\sum_{k=0}^{+\infty}\left(\frac{\partial}{c\partial t}\right)^{2k}\!\!\Delta^{-k-1}\left[\left(\frac{r}{\lambda}\right)^B \overline{N}_{n}\right]\,.
\end{equation}
This propagator depends on the length scale $\lambda$. Now our basic assumption is that in order to treat the conservative part of the dynamics, admitting the helical Killing vector $K\ua\partial\la = \partial_t + \Omega\,\partial_\varphi$ in the two-body case, one should integrate the field equations with the propagator \eqref{instpot} in which we set $\lambda=2\pi c/\Omega$. In this way the conservative dynamics will fundamentally depend on the scale $\lambda$ coming from the Killing symmetry and explicitly introduced through the propagator \eqref{instpot}.

By contrast, in a physical problem where we look for the complete dynamics including both conservative and dissipative (radiation-reaction) effects, there is no preferred scale such as $\lambda$ --- indeed, nothing suggests that the dynamics should depend on some pre-defined scale $\lambda$. In this case we integrate the field equations using the standard retarded integral, i.e.
\begin{equation}\label{retpot}
\Box_R^{-1}\left[N_{n}\right] = \mathop{\mathrm{FP}}_{B=0}\,\frac{-1}{4\pi}\int\frac{\ud^3 x'}{\vert\mathbf{x}-\mathbf{x}'\vert}\left(\frac{\vert\mathbf{x}'\vert}{\lambda}\right)^B N_{n}\left(\mathbf{x}',t-\vert\mathbf{x}-\mathbf{x}'\vert/c\right)\,.
\end{equation}
The non-linear source term $N_{n}$ is in unexpanded form since we integrate in all the exterior of the source and not only in the NZ as in \eqref{instpot}. But, as in \eqref{instpot}, we have introduced a regulator $(\vert\mathbf{x}'\vert/\lambda)^B$ and a finite part to cure the divergencies of the multipole expansion at the origin of the coordinates. Because of this regulator, the retarded integral \eqref{retpot} depends on the scale $\lambda$ which must therefore be \textit{cancelled} by other terms in the physical metric. What happens is that the dependence on $\lambda$ coming from integrating the non-linearities using \eqref{retpot} is cancelled by a related dependence on $\lambda$ of the multipole moments of the source which parametrize the linear (retarded) approximation. The source multipole moments can be written as integrals over the pseudo stress-energy tensor of the matter and gravitational fields \cite{Bl.98}. Because of the non-compactness of the gravitational field the integral extends up to infinity and involves a similar regulator $(\vert\mathbf{x}'\vert/\lambda)^B$ dealing with the boundary of the integral at infinity. The final independence of the physical metric on $\lambda$ can be checked by formally differentiating the general expression of the metric found in \cite{Bl.98}. The cancellation of $\lambda$ has been explicitly verified up to the 3PN order in the case of compact binaries \cite{Bl.al2.02}.

What is the difference between the physical situation and the ``unphysical'' one in which we would use the propagator \eqref{instpot}\,? To compare the two situations we expand the retarded integral \eqref{retpot} in the near-zone. Recalling that the overbar refers to the NZ expansion, we obtain \cite{Bl.93,PoBl.02}
\begin{equation}\label{retpotbar}
\overline{\Box_R^{-1}\left[N_{n}\right]} = \mathcal{I}^{-1}\left[\overline{N}_{n}\right] + \sum_{\ell\geqslant 0}\frac{(-)^\ell}{\ell!}\hat{\partial}_L\left\{\frac{T_L(t-r/c)-T_L(t+r/c)}{2r}\right\}\,,
\end{equation}
showing that the two solutions differ by an homogeneous solution of the wave equation which is of the anti-symmetric type (i.e. retarded minus advanced) and is therefore \textit{regular} in the source. We know that the multipolar functions $T_L(u)$ parametrizing this solution are associated with non-linear tails and their expressions can be found in \cite{Bl.93,PoBl.02}. In the physical case, the homogeneous solution in \eqref{retpotbar} will remove the $\lambda$-dependence located in the NZ logarithms appearing from the first term, and which have the symbolic NZ structure $\sim \hat{x}_L\ln(r/\lambda)$. On the other hand the $\lambda$-dependence in the FZ logarithms $\sim \hat{\partial}_L(1/r)\ln(r/\lambda)$, is removed by the retarded homogeneous solution we start with at the linear approximation. 

Now in the unphysical situation we shall want to subtract the anti-symmetric solution in \eqref{retpotbar} in order to use the instantaneous propagator $\mathcal{I}^{-1}$. Therefore the scale $\lambda$ will no longer be cancelled from the \textit{near-zone} logarithms $\sim \hat{x}_L\ln(r/\lambda)$ which will thus remain as they are. Suppose that they are evaluated at the location of a body in a two-body system, then the NZ logarithms become $\sim \hat{y}_1^L\ln(\vert\mathbf{y}_1\vert/\lambda)$ where $\mathbf{y}_1$ is the position of the body, and hence $\sim \hat{y}_1^L\ln(r_{12}/\lambda)$ in the frame of the center of mass, where $r_{12}$ is the two-body's separation. Using Kepler's law the logarithm becomes $\ln(r_{12}/\lambda)=\frac{1}{2}\ln\gamma$ where $\gamma=G m/(r_{12}c^2)$ is a standard PN parameter, showing that the NZ logarithms do contribute to the final result. 

On the contrary the FZ logarithms $\sim \hat{\partial}_L(1/r)\ln(r/\lambda)$ will not. Indeed the scale $\lambda$ therein will still be cancelled out by the linear retarded solution.\footnote{The argument could be extended to an unphysical solution which would be truly symmetric in time, i.e. which would start with a symmetric (retarded plus advanced) linear approximation and integrate the non-linearities by means of the propagator $\mathcal{I}^{-1}$.} This means that in the application to binary systems the final FZ logarithms are scaled not by $\lambda$ but rather by the size $r_{12}$ of the orbit, and become some $\sim \hat{\partial}_L(1/r)\ln(r/r_{12})$. When considered at the location of one of the bodies we get $\sim \hat{\partial}_L(1/\vert\mathbf{y}_1\vert)\ln(\vert\mathbf{y}_1\vert/r_{12})$ which clearly does not contribute in the center-of-mass frame. The latter reasoning is rather formal because the multipole expansion is valid only outside the source and it does not \textit{a priori} make sense to apply it ``at the location of one particle.'' However the reasoning may be better justified from a matching argument suggesting that the multipole expansion is valid ``everywhere'', in a restricted sense of formal asymptotic series.

Our conclusion is that we do not need to consider the FZ logarithms. From the previous investigation we see that it is sufficient to consider the family of iterated NZ logarithms generated at the quadratic order $n=2$, and to compute the 4PN and 5PN logarithms within this family. We devote the next Section to this task.

\section{The 4PN and 5PN near-zone logarithms}\label{secIII}

\subsection{External near-zone post-Newtonian metric}\label{secIIIA}

Following \cite{BlDa.88,Bl.93} we know that the dominant logarithms in the near-zone metric are coming from ``tails'' generated by quadratic coupling between the constant total mass $M$ of the system (i.e. the ADM mass) and the time varying multipole moments $M_L$ or $S_L$. Let us define $z_{1}\ab(\mathbf{n},u)$ as being the coefficient of the leading $1/r$ piece in the non-stationary or ``dynamical'' part $(h_1\ab)_\text{dyn}$ of the linearized metric given by \eqref{h1}, i.e. such that $(h_{1}\ab)_\text{dyn}=r^{-1}z_{1}\ab+\mathcal{O}(r^{-2})$. This quantity is a functional of the time varying moments (i.e. having $\ell\geqslant 2$) evaluated at retarded time $u=t-r/c$, and explicitly reads
\begin{subequations}\label{zmunu}
\begin{align}
z_{1}^{00} &= - 4 \sum_{\ell\geqslant 2}\frac{n_L}{c^{\ell+2}\ell!}M_L^{(\ell)}(u)\,,\\
z_{1}^{0i} &= - 4 \sum_{\ell\geqslant 2}\left[\frac{n_{L-1}}{c^{\ell+2}\ell!}M_{iL-1}^{(\ell)}(u) - \frac{\ell}{c^{\ell+3}(\ell+1)!}\varepsilon_{iab}\,n_{aL-1}S_{bL-1}^{(\ell)}(u)\right]\,,\\
z_{1}^{ij} &= - 4 \sum_{\ell\geqslant 2}\left[\frac{n_{L-2}}{c^{\ell+2}\ell!}M_{ijL-2}^{(\ell)}(u) -\frac{2\ell}{c^{\ell+3}(\ell+1)!}n_{aL-2}\,\varepsilon_{ab(i}S_{j)bL-2}^{(\ell)}(u)\right]\,.
\end{align}\end{subequations}
All the logarithms in the quadratic metric $h_{2}\ab$ will be generated from the leading $1/r^2$ piece in the quadratic source, defined by $N_{2}\ab=r^{-2}Q_{2}\ab(\mathbf{n},u)+\mathcal{O}(r^{-3})$. The coefficient is computed from the quantity \eqref{zmunu} as $Q_{2}\ab = \frac{4 M}{c^4}\,{}^{(2)}z_{1}\ab + \frac{k^\alpha k^\beta}{c^2}\sigma$, where the first term will generate the tails, and the second term is associated with the stress-energy of gravitational waves, with $k^\alpha=(1,\mathbf{n})$ the Minkowskian outgoing null vector, and $\sigma=\frac{1}{2}{}^{(1)}{z_{1}}^{\mu\nu}{}^{(1)}{z_{1}}_{\mu\nu}-\frac{1}{4}{}^{(1)}{z_{1}}^{\mu}_{\mu}{}^{(1)}{z_{1}}^{\nu}_{\nu}$. Now, as shown in Appendix \ref{appA}, the logarithms produced by the second term $\propto k^\alpha k^\beta$ are pure gauge, so only the first term dealing with tails is responsible for the near-zone logarithms. Hence the part of the NZ expansion of the quadratic metric $\overline{h}_{2}$ containing those logarithms is given by
\begin{equation}\label{deltah2}
\delta\overline{h}_{2}\ab = \mathop{\mathrm{FP}}_{B=0}\,\sum_{k=0}^{+\infty}\left(\frac{\partial}{c\partial t}\right)^{2k}\!\!\Delta^{-k-1}\left[\left(\frac{r}{\lambda}\right)^B\frac{4 M}{r^2 c^4} \mathop{z_{1}}^{(2)}\!{}\ab(\mathbf{n},u)\right]\,.
\end{equation}
We substitute the explicit expression \eqref{zmunu} into \eqref{deltah2}, expand the retardation $u=t-r/c$ in the source term when $r\rightarrow 0$, and integrate using Eqs.~\eqref{mathieu}--\eqref{alphaB}. Then we look for the poles $\propto 1/B$ and after applying the finite part get the logarithms. Some general formulas for obtaining the logarithms directly from the unexpanded source are relegated to Appendix \ref{appA}. We readily recover that the dominant logarithms arise at 4PN order. We limit our computation to the leading order 4PN and next-to-leading order 5PN logarithms, and find
\begin{subequations}\label{deltah2exp}
\begin{align}
\delta\overline{h}_{2}^{00} &= - \frac{8 M}{15 c^{10}}\left\{x^{ab}\left[M^{(6)}_{ab}+\frac{1}{14}\frac{r^2}{c^2}M^{(8)}_{ab}\right]-\frac{1}{21}\frac{x^{abc}}{c^2}M^{(8)}_{abc}\right\}\ln\left(\frac{r}{\lambda}\right)+\mathcal{O}\left(\frac{1}{c^{14}}\right)\,,\\
\delta\overline{h}_{2}^{0i} &= \frac{8 M}{3 c^{9}}\left\{x^{a}\left[M^{(5)}_{ai}+\frac{1}{10}\frac{r^2}{c^2}M^{(7)}_{ai}\right]-\frac{1}{15}\frac{x^{ab}}{c^2}M^{(7)}_{abi}+\frac{2}{15}\varepsilon_{iab}\frac{x^{ac}}{c^2}S^{(6)}_{bc}\right\}\ln\left(\frac{r}{\lambda}\right)+\mathcal{O}\left(\frac{1}{c^{13}}\right)\,,\\
\delta\overline{h}_{2}^{ij} &= -\frac{8 M}{c^{8}}\left\{M^{(4)}_{ij}+\frac{1}{6}\frac{r^2}{c^2}M^{(6)}_{ij}-\frac{1}{9}\frac{x^{a}}{c^2}M^{(6)}_{aij}+\frac{4}{9}\frac{x^{a}}{c^2}\varepsilon_{ab(i}S^{(5)}_{j)b}\right\}\ln\left(\frac{r}{\lambda}\right)+\mathcal{O}\left(\frac{1}{c^{12}}\right)\,.
\end{align}\end{subequations}
The mass-type quadrupole moment $M_{ij}$, mass octupole moment $M_{ijk}$ and current quadrupole $S_{ij}$ in Eqs.~\eqref{deltah2exp} are functions of coordinate time $t$. The indicated PN remainders $\mathcal{O}(c^{-p})$ refer only to the logarithmic terms.

We now want to iterate the expressions \eqref{deltah2exp} at higher non-linear order in order to get the complete family of logarithms generated by that ``seed''. To do that it is very convenient to perform first a change of gauge. Starting from \eqref{deltah2exp}, which is defined in some harmonic gauge, we pose $\overline{k}\ab_{2}=\overline{h}\ab_{2}+2\partial^{(\alpha}\xi_{2}^{\beta)}-\eta\ab\partial_\mu\xi_{2}^\mu$ with gauge vector
\begin{subequations}\label{xi2}
\begin{align}
\xi_{2}^{0} &= \frac{M}{c^{9}}\left\{\frac{2}{3}x^{ab}M^{(5)}_{ab}+\frac{1}{21}\frac{r^2}{c^2}x^{ab}M^{(7)}_{ab}-\frac{2}{135}\frac{x^{abc}}{c^2}M^{(7)}_{abc}\right\}\ln\left(\frac{r}{\lambda}\right)\,,\\
\xi_{2}^{i} &= \frac{M}{c^{8}}\left\{4x^aM^{(4)}_{ai}+\frac{2}{3}\frac{r^2}{c^2}x^{a}M^{(6)}_{ai}-\frac{2}{3}\frac{x^{iab}}{c^2}M^{(6)}_{ab}-\frac{2}{9}\frac{x^{ab}}{c^2}M^{(6)}_{abi}+\frac{16}{9}\varepsilon_{iab}\frac{x^{ac}}{c^2}S^{(5)}_{bc}\right\}\ln\left(\frac{r}{\lambda}\right)\,.
\end{align}\end{subequations}
This gauge transformation will have the effect of moving many 4PN logarithmic terms into the 00 component of the (ordinary covariant) metric. As a result the implementation of the non-linear iteration in that new gauge will be especially simple. Since our aim is to compute the gauge invariant quantity \eqref{uT} we can work in any convenient gauge. Our chosen gauge is very similar to the generalization of the Burke-Thorne gauge introduced in \cite{Bl.97} to deal with higher-order (2.5PN and 3.5PN) radiation-reaction effects. We obtain
\begin{subequations}\label{deltak2}
\begin{align}
&\delta\overline{k}_{2}^{00}+\delta\overline{k}_{2}^{ii} =  \frac{M}{c^{10}}\left\{-\frac{16}{5}x^{ab}M^{(6)}_{ab}-\frac{8}{35}\frac{r^2}{c^2}x^{ab}M^{(8)}_{ab}+\frac{16}{189}\frac{x^{abc}}{c^2}M^{(8)}_{abc}\right\}\ln\left(\frac{r}{\lambda}\right)+\mathcal{O}\left(\frac{1}{c^{14}}\right)\,,\label{deltak200}\\
&\delta\overline{k}_{2}^{0i} = \frac{M}{c^{11}}\left\{\frac{16}{21}\hat{x}^{iab}M^{(7)}_{ab}-\frac{64}{45}\varepsilon_{iab}x^{ac}S^{(6)}_{bc}\right\}\ln\left(\frac{r}{\lambda}\right)+\mathcal{O}\left(\frac{1}{c^{13}}\right)\,,\\
&\delta\overline{k}_{2}^{ij} = \mathcal{O}\left(\frac{1}{c^{12}}\right)\,.
\end{align}\end{subequations}
In this gauge the iteration at cubic non-linear order is very simple. To control all the 5PN logarithmic terms at cubic order $n=3$ we need only to solve the Poisson equation $\Delta[\delta\overline{k}_{3}^{00}+\delta\overline{k}_{3}^{ii}] = - 2 \partial_j\overline{h}_{1}^{00}\,\partial_j\delta\overline{k}_{2}^{00} + \mathcal{O}(c^{-14})$, where $\overline{h}_{1}^{00}$ denotes the NZ expansion of the linearized metric \eqref{h100}, and we can use for $\delta\overline{k}_{2}^{00}$ the leading 4PN approximation given by the first term in \eqref{deltak200}. Posing $\overline{h}_{1}^{00}=-4\overline{U}/c^2+\mathcal{O}(c^{-4})$, the latter equation is integrated as\footnote{Actually the integration yields in addition to the near-zone 5PN logarithm  \eqref{deltak3} the extra far-zone 5PN logarithmic contribution 
$$\left(\delta\overline{k}_{3}^{00}+\delta\overline{k}_{3}^{ii}\right)_\text{FZ} = -\frac{64M}{c^{12}} M_{ab}^{(6)}\ln\left(\frac{r}{\lambda}\right)\sum_{\ell\geqslant0}\frac{(-)^\ell}{(2\ell+5)\ell!}\partial_L\left(\frac{1}{r}\right)M_{Lab}\,.$$
We argued on general grounds in Sec.~\ref{secIIC} that FZ logarithms do not have to be considered for the present computation, so we drop this term out in the following.}
\begin{equation}\label{deltak3}
\delta\overline{k}_{3}^{00}+\delta\overline{k}_{3}^{ii} = - \frac{64M}{5c^{12}}\overline{U} x^{ab}M^{(6)}_{ab}\ln\left(\frac{r}{\lambda}\right)+ \mathcal{O}\left(\frac{1}{c^{14}}\right)\,,
\end{equation}
with the explicit expression
\begin{equation}\label{Ubar}
\overline{U} = \sum_{\ell=0}^{+\infty}\frac{(-)^\ell}{\ell!}M_L\,\partial_L\left(\frac{1}{r}\right)\,.
\end{equation}
We readily check that the quartic and higher non-linear iterations ($n\geqslant 4$) are not needed for controlling the 4PN and 5PN logarithmic terms (cf. the discussion at the end of Sec.~\ref{secIIB}).

\subsection{Internal near-zone post-Newtonian metric}\label{secIIIB}

The metric we computed so far is in the form of a multipolar expansion valid in the exterior of an isolated source. We now want to deduce from it the metric \textit{inside} the matter source. First of all, since the expressions \eqref{deltak2} are regular at the origin $r\rightarrow 0$, we find using a matching argument that they are necessarily also valid inside the matter source. On the other hand it is clear that the expression \eqref{deltak3} will also be valid inside the source provided that we match the multipole expansion $\overline{U}$ given by \eqref{Ubar} with the actual Newtonian potential of the source. From the known Newtonian limit of the multipole moments $M_L=\int\ud^3x\,\hat{x}^L\rho(\mathbf{x},t)+\mathcal{O}(c^{-2})$, where $\rho$ is the Newtonian source density in the source, we get $G\overline{U}=U+\mathcal{O}(c^{-2})$ where
\begin{align}\label{Upoisson}
U = G \int\frac{\ud^3x'}{\vert\mathbf{x}-\mathbf{x}'\vert}\,\rho(\mathbf{x}',t)\,. 
\end{align}
From the latter arguments we therefore obtain the piece of the inner metric of any isolated source (coming back to the usual covariant metric $g_{\alpha\beta}$) that depends logarithmically on the distance $r$ to the source's center at 4PN and 5PN orders as
\begin{subequations}\label{deltag'}
\begin{align}
\delta'{g}_{00} &= \frac{G^2M}{c^{10}}\left[\frac{8}{5}\left(1-\frac{2U}{c^2}\right)x^{ab}M^{(6)}_{ab}+\frac{4}{35c^2}r^2x^{ab}M^{(8)}_{ab}-\frac{8}{189c^2}x^{abc}M^{(8)}_{abc}\right]\ln\left(\frac{r}{\lambda}\right)+ \mathcal{O}\left(\frac{1}{c^{14}}\right)\,,\label{deltag00}\\
\delta'{g}_{0i} &= \frac{G^2M}{c^{11}}\left[\frac{16}{21}\hat{x}^{iab}M^{(7)}_{ab}-\frac{64}{45}\varepsilon_{iab}x^{ac}S^{(6)}_{bc}\right]\ln\left(\frac{r}{\lambda}\right) + \mathcal{O}\left(\frac{1}{c^{13}}\right)\,,\\
\delta'{g}_{ij} &= \frac{G^2M}{c^{10}}\left[\frac{8}{5}x^{ab}M^{(6)}_{ab}\delta_{ij}\right]\ln\left(\frac{r}{\lambda}\right) + \mathcal{O}\left(\frac{1}{c^{12}}\right)\,,
\end{align}\end{subequations}
where $U$ is the Newtonian potential \eqref{Upoisson} valid all over the source. 

However we now discuss other pieces of the inner metric whose near-zone expansion does not explicitly depend on the logarithms of $r$ but which involve new inner potentials integrating over a logarithmically modified source density. The first of these pieces comes from the fact that the 4PN modification of the metric given by the first term in \eqref{deltag00} implies a modification of the stress-energy tensor of the matter fluid at the 5PN order; in particular the fluid's source density, say $\sigma=T^{00}/c^2$, gets modified by the amount
\begin{equation}\label{deltarho}
\frac{\delta\sigma}{\rho} = -\frac{4}{5}\frac{G^2M}{c^{10}}x^{ab}M^{(6)}_{ab}\ln\left(\frac{r}{\lambda}\right)+ \mathcal{O}\left(\frac{1}{c^{12}}\right)\,.
\end{equation}
On the other hand the 4PN term of the metric will induce a 4PN change in the acceleration of the fluid motion given by
\begin{equation}\label{deltaacc}
\delta a^i = \frac{8}{5}\frac{G^2M}{c^{8}}x^{a}M^{(6)}_{ai}\ln\left(\frac{r}{\lambda}\right)+ \mathcal{O}\left(\frac{1}{c^{10}}\right)\,. 
\end{equation}
When computing the inner metric at the 1PN order we have to take into account the retardation due to the propagation of gravity, using say $\Box^{-1}=\Delta^{-1}+c^{-2}\partial_t^2 \Delta^{-2}+\mathcal{O}(c^{-4})$. The time derivatives at 1PN order will yield an acceleration and the modification of the acceleration \eqref{deltaacc} will give a contribution at 5PN order. We find that the sum of the two effects gives the following extra contribution to the inner metric at 5PN order
\begin{equation}\label{deltag''}
\delta''{g}_{00} = -\frac{8}{5}\frac{G^3M}{c^{12}}x^{a}M^{(6)}_{ab}\int\frac{\ud^3x'}{\vert\mathbf{x}-\mathbf{x}'\vert}\,\rho'\,x'^{b}\ln\left(\frac{r'}{\lambda}\right)+ \mathcal{O}\left(\frac{1}{c^{14}}\right)\,.
\end{equation}
This 5PN contribution is present only in the 00 component of the metric.\footnote{Interestingly, it was found in Ref.~\cite{IyWi.95} (following \cite{IyWi.93}) that a similar looking contribution must also be taken into account when computing the higher-order (3.5PN) radiation-reaction force for compact binary systems from a near-zone radiation-reaction formalism defined in \cite{Bl.97}. Actually the 2.5PN+3.5PN near-zone radiation-reaction formalism \cite{Bl.97} (see in particular Eqs.~(2.16) there) is quite similar to the present 4PN+5PN near-zone conservative logarithm formalism.} The complete logarithmic contributions we shall consider in this paper are thus given by 
\begin{equation}\label{deltag}
\delta{g}\lab = \delta'{g}\lab + \delta''{g}\lab\,.
\end{equation}
These contributions exhaust the possibilities of having 4PN and 5PN near-zone logarithmic terms in the gauge invariant observable quantity \eqref{uT}.

\subsection{Application to compact binary systems}\label{secIIIC}

Let us now apply the previous results to the specific problem of a system of two point particles. The Newtonian mass density in that case is $\rho = \sum_a m_a\delta(\mathbf{x}-\mathbf{y}_a)$ where $\delta$ is the Dirac delta function. The trajectory of the $a$-th particle ($a$=1,2) is denoted $\mathbf{y}_a(t)$; the ordinary coordinate velocity will be $\mathbf{v}_a=\ud \mathbf{y}_a/\ud t$. The two masses $m_a$ have sum $m=m_1+m_2$, reduced mass $\mu=m_1m_2/m$ and symmetric mass ratio $\nu=\mu/m$. The Newtonian potential of the system reduces to 
\begin{equation}\label{U}
U = \frac{G m_1}{r_1} +  \frac{G m_2}{r_2}\,,
\end{equation}
where $r_a=\vert\mathbf{x}-\mathbf{y}_a\vert$ is the distance from particle $a$. The regularized value of that potential at the location of particle 1 is simply
\begin{equation}\label{U1}
U_1 = \frac{G m_2}{r_{12}}\,,
\end{equation}
where $r_{12}=\vert\mathbf{y}_1-\mathbf{y}_2\vert$. Similarly we evaluate the logarithmic contributions at the location of particle 1. Concerning the first piece \eqref{deltag'} we find (no longer mentioning the PN remainder)
\begin{subequations}\label{deltag'1}
\begin{align}
\delta'{g}_{00}(\mathbf{y}_1) &= \frac{G^2M}{c^{10}}\left[\frac{8}{5}\left(1-\frac{2U_1}{c^2}\right)y_1^{ab}M^{(6)}_{ab}+\frac{4}{35c^2}y_1^2y_1^{ab}M^{(8)}_{ab}-\frac{8}{189c^2}y_1^{abc}M^{(8)}_{abc}\right]\ln\left(\frac{\vert\mathbf{y}_1\vert}{\lambda}\right)\,,\\
\delta'{g}_{0i}(\mathbf{y}_1) &= \frac{G^2M}{c^{11}}\left[\frac{16}{21}\hat{y}_1^{iab}M^{(7)}_{ab}-\frac{64}{45}\varepsilon_{iab}y_1^{ac}S^{(6)}_{bc}\right]\ln\left(\frac{\vert\mathbf{y}_1\vert}{\lambda}\right)\,,\\
\delta'{g}_{ij}(\mathbf{y}_1) &= \frac{G^2M}{c^{10}}\left[\frac{8}{5}y_1^{ab}M^{(6)}_{ab}\delta_{ij}\right]\ln\left(\frac{\vert\mathbf{y}_1\vert}{\lambda}\right) \,,
\end{align}\end{subequations}
which involves the logarithm $\ln(r/\lambda)$ evaluated on the particle 1, i.e. $\ln(\vert\mathbf{y}_1\vert/\lambda)$. As for the second piece \eqref{deltag''} we compute the Poisson integral using $\rho = \sum_a m_a\delta(\mathbf{x}-\mathbf{y}_a)$ and perform a regularization on the particle 1 to obtain
\begin{equation}\label{deltag''1}
\delta''{g}_{00}(\mathbf{y}_1) = -\frac{8}{5}\frac{G^2M}{c^{12}}\,U_1\,y_1^{a}y_2^{b}M^{(6)}_{ab}\,\ln\left(\frac{\vert\mathbf{y}_2\vert}{\lambda}\right)\,,
\end{equation}
which is proportional to the logarithm $\ln(\vert\mathbf{y}_2\vert/\lambda)$ associated with the other particle. These results are valid in a general frame. However we shall later specify the origin of the coordinate system to be the center of mass of the binary system. In that case we have $\ln(\vert\mathbf{y}_a\vert/\lambda)=\ln(r_{12}/\lambda)+\ln(\mu/m_a)+\mathcal{O}(c^{-2})$, where the PN remainder does not involve any logarithmic term, and the logarithm of the mass ratio is a constant, and is therefore clearly irrelevant to our search of logarithmic terms; so $\ln(r_{12}/\lambda)$ is in fact the only relevant logarithm and we shall now systematically replace all $\ln(\vert\mathbf{y}_a\vert/\lambda)$'s by $\ln(r_{12}/\lambda)$. Finally we end up with the following contributions of the 4PN and 5PN logarithms to the near-zone metric evaluated at the location of particle 1 in our chosen gauge,
\begin{subequations}\label{deltag1}
\begin{align}
\delta{g}_{00}(\mathbf{y}_1) &= \frac{G^2M}{c^{10}}\left[\frac{8}{5}\left(1-\frac{2U_1}{c^2}\right)y_1^{ab}M^{(6)}_{ab}-\frac{8}{5c^2}U_1y_1^{a}y_2^{b}M^{(6)}_{ab}\right.\nonumber\\&\left.\qquad\quad+\frac{4}{35c^2}y_1^2y_1^{ab}M^{(8)}_{ab}-\frac{8}{189c^2}y_1^{abc}M^{(8)}_{abc}\right]\ln\left(\frac{r_{12}}{\lambda}\right)\,,\\
\delta{g}_{0i}(\mathbf{y}_1) &= \frac{G^2M}{c^{11}}\left[\frac{16}{21}\hat{y}_1^{iab}M^{(7)}_{ab}-\frac{64}{45}\varepsilon_{iab}\,y_1^{ac}S^{(6)}_{bc}\right]\ln\left(\frac{r_{12}}{\lambda}\right)\,,\\
\delta{g}_{ij}(\mathbf{y}_1) &= \frac{G^2M}{c^{10}}\left[\frac{8}{5}y_1^{ab}M^{(6)}_{ab}\delta_{ij}\right]\ln\left(\frac{r_{12}}{\lambda}\right)\,.
\end{align}\end{subequations}
Note that this result is complete but not fully explicit because we have still to replace all the multipole moments $M_L$ and $S_L$ by their expressions valid for point mass binary systems. In particular the quadrupole mass moment $M_{ij}$ should be given with 1PN relative precision as ($1\leftrightarrow 2$ means adding the same terms for particle 2)
\begin{equation}\label{Mij1PN}
M_{ij} = m_1\left\{\left[1+\frac{1}{c^2}\left(\frac{3}{2}v_1^2-\frac{G m_2}{r_{12}}\right)\right]\hat{y}_1^{ij}+\frac{1}{14c^2}\frac{\ud^2}{\ud t^2}\left(y_1^2\hat{y}_1^{ij}\right)-\frac{20}{21c^2}\frac{\ud}{\ud t}\left(v_1^k\hat{y}_1^{ijk}\right)\right\}+ 1\leftrightarrow 2\,,
\end{equation}
and its time derivatives should consistently use the 1PN equations of motion. Besides $M_{ij}$ we also need the constant mass monopole or total mass $M$ at 1PN order, namely
\begin{equation}\label{M1PN}
M = m_1\left[1+\frac{1}{c^2}\left(\frac{1}{2}v_1^2-\frac{G m_2}{2r_{12}}\right)\right]+ 1\leftrightarrow 2\,.
\end{equation}
All the other moments are only required at the Newtonian accuracy, and read
\begin{subequations}\label{MLSL}
\begin{align}
M_{L} &= m_1\,\hat{y}_1^L + 1\leftrightarrow 2\,,\\
S_L &= m_1\,\varepsilon^{ab\langle i_\ell} \,y_1^{L-1\rangle a}v_1^b + 1\leftrightarrow 2\,.
\end{align}\end{subequations}
However in applications it is often better to postpone the (messy) replacements of the multipole moments by their explicit values \eqref{Mij1PN}--\eqref{MLSL} and to use more compact expressions such as \eqref{deltag1}.

\section{Logarithms in the equations of motion and energy}\label{secIV}

\subsection{General orbits}\label{secIVA}

With the 4PN and 5PN logarithmic contributions in the near-zone metric \eqref{deltag1} we now derive the corresponding terms in the acceleration of point particle binary systems. The computation is straightforward from the geodesic equation. A subtle point is that we must take into account the coupling between the 1PN terms in the metric and the 4PN logarithm to produce new 5PN logarithms. On the other hand one must be careful about the replacement of accelerations in 1PN terms by the 4PN acceleration to also produce 5PN logarithms. The final result, valid for generic (non-circular) orbits in an arbitrary frame, is
\begin{align}\label{deltaa1}
\delta a_1^i &= \frac{G^2M}{c^8}\left[\frac{8}{5}y_1^aM^{(6)}_{ia}\right.\nonumber\\
&\quad + \frac{1}{c^2}\left( \frac{8}{5}v_1^2y_1^aM^{(6)}_{ia}-\frac{32}{5}U_1y_1^aM^{(6)}_{ia}+\frac{28}{5}U_1y_2^aM^{(6)}_{ia}+\frac{16}{5}U_1r_{12}^{-2}y_{12}^iy_1^{ab}M^{(6)}_{ab}\right.\nonumber\\
&\quad\quad+\frac{4}{5}U_1r_{12}^{-2}y_{12}^iy_1^{a}y_2^{b}M^{(6)}_{ab}-\frac{68}{105}y_1^{iab}M^{(8)}_{ab}+\frac{44}{105}y_1^2y_1^{a}M^{(8)}_{ia}-\frac{4}{63}y_1^{ab}M^{(8)}_{iab}\nonumber\\
&\quad\quad-\frac{32}{5}v_1^{ia}y_1^{b}M^{(6)}_{ab}-\frac{12}{5}v_1^{i}y_1^{ab}M^{(7)}_{ab}-\frac{32}{15}y_1^{ia}v_1^{b}M^{(7)}_{ab}+\frac{32}{15}(y_1v_1)y_1^{a}M^{(7)}_{ia}\nonumber\\
& \quad\quad\left.\left.+\frac{128}{45}\varepsilon_{iab}v_1^{a}y_1^{c}S^{(6)}_{bc}+\frac{64}{45}\varepsilon_{iab}y_1^{a}v_1^{c}S^{(6)}_{bc}-\frac{64}{45}\varepsilon_{abc}y_1^{a}v_1^{c}S^{(6)}_{ib}+\frac{64}{45}\varepsilon_{iab}y_1^{ac}S^{(7)}_{bc}\right)\right]\ln\left(\frac{r_{12}}{\lambda}\right)\,,
\end{align}
where the multipole moments are given by \eqref{Mij1PN}--\eqref{MLSL}.

An important check of this result is that the acceleration should be purely \textit{conservative}, by which we mean that there should exist some corresponding contributions at the 4PN and 5PN orders in the conserved energy, angular momentum, linear momentum and center-of-mass position of the binary system. Let us see how this works in the case of the energy. The modification at 4PN and 5PN of the energy, say $\delta E$, should be such that $\ud \delta E/\ud t$ exactly balances the replacement of accelerations by \eqref{deltaa1} in the time derivative of the known expression of the energy up to 1PN order (say $E_\text{1PN}$). This requirement yields
\begin{equation}\label{dtdeltaE}
\frac{\ud \delta E}{\ud t} = - m_1\left[v_1^i + \frac{1}{c^2}\left(\frac{3}{2} v_1^2 v_1^i +\frac{G m_2}{r_{12}}\left[-\frac{1}{2}(n_{12}v_2) n_{12}^i + 3v_{1}^i - \frac{7}{2}v_{2}^i\right]\right)\right]\delta a_1^i + 1 \leftrightarrow 2\,.
\end{equation}
Plugging \eqref{deltaa1} into \eqref{dtdeltaE} and using the expressions of the multipole moments \eqref{Mij1PN}--\eqref{MLSL}, we indeed find that the right-hand-side of \eqref{dtdeltaE} takes the form of a total time derivative, and we are thus able to infer the contribution to the energy,
\begin{align}\label{deltaE}
\delta E &=  \frac{G^2M}{c^8}\left[-\frac{4}{5}M^{(5)}_{ab}M^{(1)}_{ab}+\frac{4}{5}M^{(4)}_{ab}M^{(2)}_{ab}-\frac{2}{5}M^{(3)}_{ab}M^{(3)}_{ab}\right.\nonumber\\
&\quad\quad + \frac{1}{c^2}\left(\frac{4}{189}M^{(7)}_{abc}M^{(1)}_{abc}-\frac{4}{189}M^{(6)}_{abc}M^{(2)}_{abc}+\frac{4}{189}M^{(5)}_{abc}M^{(3)}_{abc}-\frac{2}{189}M^{(4)}_{abc}M^{(4)}_{abc}\right.\nonumber\\
&\quad \quad\quad+\frac{64}{45}S^{(6)}_{ab}S_{ab}-\frac{64}{45}S^{(5)}_{ab}S^{(1)}_{ab}+\frac{64}{45}S^{(4)}_{ab}S^{(2)}_{ab}-\frac{32}{45}S^{(3)}_{ab}S^{(3)}_{ab}+\frac{12}{5}M^{(6)}_{ab}Q_{ab}\nonumber\\
&\quad \quad\quad\left.\left.+\frac{2}{35}M^{(6)}_{ab}H^{(2)}_{ab}-\frac{2}{35}M^{(7)}_{ab}H^{(1)}_{ab}-\frac{16}{21}M^{(6)}_{ab}K^{(1)}_{ab}+\frac{16}{21}M^{(7)}_{ab}K_{ab}\right)\right]\ln\left(\frac{r_{12}}{\lambda}\right)\,.
\end{align}
In addition to the standard multipole moments \eqref{Mij1PN}--\eqref{MLSL}, we have also introduced the supplementary moments (needed only at Newtonian accuracy)
\begin{subequations}\label{HKQ}
\begin{align}
H_L &= m_1 y_1^2 \,\hat{y}_1^L + 1\leftrightarrow 2\,,\\
K_L &= m_1 v_1^i \,\hat{y}_1^{iL} + 1\leftrightarrow 2\,,\\
Q_L &= m_1 v_1^2 \,\hat{y}_1^L + 1\leftrightarrow 2\,.
\end{align}\end{subequations}
By the same method we have also computed the modification of another integral of the motion, namely the center-of-mass position $G^i$. The result is
\begin{align}\label{deltaGi}
\delta G^i &=  \left[ \frac{2GM}{c^3} M_i^{(1)} + \frac{G^2M}{c^8} \left( \frac{16}{5}M^{(1)}_{a}M^{(3)}_{ai}-\frac{8}{5}M_{a}M^{(4)}_{ai}\right)\right.\nonumber\\ &\quad\quad + \frac{G^2 M}{c^{10}}\left(\frac{68}{105}M^{(6)}_{ab}M_{iab}-\frac{20}{63}M^{(5)}_{ab}M^{(1)}_{iab}+\frac{16}{63}M^{(4)}_{ab}M^{(2)}_{iab}-\frac{4}{21}M^{(3)}_{ab}M^{(3)}_{iab}\right.\nonumber\\ &\quad \quad\quad +\frac{4}{21}M^{(2)}_{ab}M^{(4)}_{iab}-\frac{8}{63}M^{(1)}_{ab}M^{(5)}_{iab}+\frac{4}{63}M_{ab}M^{(6)}_{iab}-\frac{64}{45}\varepsilon_{iab}M_{ac}S^{(5)}_{bc}\nonumber\\ &\quad \quad\quad+\frac{32}{45}\varepsilon_{iab}M^{(1)}_{ac}S^{(4)}_{bc}+\frac{32}{45}\varepsilon_{iab}M^{(4)}_{ac}S^{(1)}_{bc}-\frac{64}{45}\varepsilon_{iab}M^{(5)}_{ac}S_{bc}\nonumber\\ &\quad \quad\quad\left.\left.-\frac{32}{15}S_{a}S^{(4)}_{ia}-\frac{4}{25}H_{a}M^{(6)}_{ia}\right)\right]\ln\left(\frac{r_{12}}{\lambda}\right)\,,
\end{align}
and will be useful when restricting our general result for the redshift observable --- valid for a generic orbit in an arbitrary frame --- to circular orbits described in the center-of-mass frame.

\subsection{Circular orbits}

Let us now focus our attention on the case of circular orbits. We look for the 4PN and 5PN logarithms $\delta a_{12}^i$ in the relative acceleration $a_{12}^i=a_1^i-a_2^i$ of the particles for circular orbits. The first contribution to $\delta a_{12}^i$ will evidently come from the difference $\delta a_1^i-\delta a_2^i$. We insert the center-of-mass relations $y_a^i=Y_a^i[\mathbf{y}_{12},\mathbf{v}_{12}]$, expressing the individual positions in terms of the relative position $y_{12}^i=y_1^i-y_2^i$ and relative velocity $v_{12}^i=v_1^i-v_2^i=\ud y_{12}^i/\ud t$. At 1PN order and for circular orbits these expressions simply reduce to the Newtonian relations $y_1^i=X_2 \, y_{12}^i$ and $y_2^i=-X_1 \, y_{12}^i$, where $X_a=m_a/m$. All the multipole moments and their time derivatives are replaced by their expressions for circular orbits given in terms of $y_{12}^i$ and $v_{12}^i$ (and masses). However there is also another contribution which comes from the known relative acceleration at 1PN order (say $a_\text{1PN}^i$) when reduced to circular orbits. As usual we perform an iterative computation: knowing first $\delta a_{12}^i$ at 4PN order we use the result to find the next order 5PN correction. In this computation we use the fact that the center-of-mass relations $y_a^i=Y_a^i[\mathbf{y}_{12},\mathbf{v}_{12}]$ are not modified by logarithmic terms before the 5PN order. This is checked using the modification of the integral of the center-of-mass $\delta G^i$ given in \eqref{deltaGi} (see also the result \eqref{deltaGic} for circular orbits below, which is clearly a 5PN effect). Finally the modification of the acceleration is found to be of the form
\begin{equation}\label{deltaa}
\delta a_{12}^i = - \delta\Omega^2 \,y_{12}^i\,,
\end{equation}
where the total change in the orbital frequency (squared) for circular orbits due to 4PN and 5PN logarithms reads
\begin{equation}\label{deltaOm2}
    \delta\Omega^2 = \frac{G m}{r_{12}^3} \left[ \frac{128}{5} + \left(-\frac{8572}{35}-112\nu\right)\gamma \right] \nu\,\gamma^4\,\ln\gamma \,.
\end{equation}
The orbital separation is $r_{12}=\vert\mathbf{y}_{12}\vert$, and we have introduced the convenient post-Newtonian parameter (where $m=m_1+m_2$)
\begin{equation}\label{gamma}
\gamma = \frac{G m}{r_{12} c^2}\,.
\end{equation}
From \eqref{deltaOm2} we have the relation between the orbital frequency and the parameter $\gamma$. Inverting this relation we obtain $\gamma$ as a function of the orbital frequency or, rather, of the parameter $x$ defined by
\begin{equation}\label{x}
x = \left(\frac{G \, m \, \Omega}{c^3}\right)^{2/3}\,.
\end{equation}
We find that the 4PN and 5PN logarithms in $\gamma$ as a function of $x$ are
\begin{align}\label{deltagam}
    \delta\gamma = \left[-\frac{128}{15} + \left(\frac{508}{105}+\frac{944}{15}\nu\right)x\right] \nu\,x^5\,\ln x\,.
\end{align}

We have taken into account in \eqref{deltaOm2} and \eqref{deltagam} the important fact that the length scale $\lambda=cP$ is related to the period $P=2\pi/\Omega$, and hence contributes to the logarithm. As already mentioned, using Kepler's third law we have $\frac{r_{12}}{\lambda}=\frac{\sqrt{\gamma}}{2\pi}$, so that $\ln(\frac{r_{12}}{\lambda}) = \frac{1}{2}\ln\gamma$ plus an irrelevant constant. Post-Newtonian corrections to Kepler's law do not change the argument, which applies with $x$ as well as with $\gamma$. Recall that $\lambda=2\pi c/\Omega$ was introduced in the problem when we assumed the existence of the helical Killing vector $K\ua\partial\la = \partial_t + \Omega\,\partial_\varphi$ to describe exactly circular orbits. Then this scale entered explicitly into the propagator we used to integrate the field [see \eqref{hexp} or \eqref{deltah2}], and it is thus no surprise that it contributes to the final result. Of course we could have chosen any other scale proportional to $\lambda$ without changing the result which concerns only the logarithmic dependence.

To be clearer about formulas such as \eqref{deltaOm2} and \eqref{deltagam} we would need to give the more complete formulas including also the known contributions up to 3PN order. However we must be careful since these formulas depend on the gauge. Thus $\delta\Omega^2$ and $\delta\gamma$ are to be added to the 3PN expressions given by Eqs.~(188) and (191) in \cite{Bl.06} when working in Hadamard regularization gauge, or by Eqs.~(B6) and (B7) in Paper I when working in dimensional regularization gauge. Also the 4PN and 5PN terms computed in \eqref{deltaOm2} and \eqref{deltagam} themselves depend on the choice of gauge at the 4PN and 5PN orders (see Sec.~\ref{secIII}).

It is much better to turn to gauge invariant quantities. The most obvious one is the conserved energy $E$ for circular orbits as a function of the orbital frequency $\Omega$. As for the previous computation of the acceleration we have two contributions, one coming directly from the general-orbit modification of the energy given by \eqref{deltaE}, and one coming from the circular-orbit reduction of the 1PN energy $E_\text{1PN}$. We first express the results entirely in terms of the parameter $\gamma$ using \eqref{deltaOm2} and then replace all the $\gamma$'s by functions of the $x$'s using \eqref{deltagam}. The result for the 4PN and 5PN logarithms is (where $\mu = m_1 m_2 / m$)
\begin{equation}\label{deltaEx}
    \delta E = -\frac{1}{2} \mu \, c^2 x \left[ \frac{448}{15} + \left(-\frac{4988}{35}-\frac{1904}{15}\nu\right)x \right] \nu\,x^4\,\ln x \,.
\end{equation}
Beware that $\delta E$ here has not the same meaning as in \eqref{deltaE} because of the additional terms coming from the circular-orbit reduction of the 1PN energy $E_\text{1PN}$.

Since the energy as a function of $x$ is a gauge invariant relation, let us also provide the complete result including all the known terms up to 3PN order, and also the 4PN and 5PN terms in the test-mass limit for one of the particles known from the exact result $\lim_{\nu\rightarrow 0}E/(\mu c^2)=(1-2x)(1-3x)^{-1/2}-1$. We have
\begin{align}\label{Ecomplet}
    E =& -\frac{1}{2} \mu \, c^2 x  \left\{ 1 + \left( -\frac{3}{4} - \frac{\nu}{12} \right) x + \left( - \frac{27}{8} + \frac{19}{8} \nu - \frac{\nu^2}{24} \right) x^2 \right. \nonumber \\ &\qquad+ \left. \left( - \frac{675}{64} + \left[ \frac{34445}{576} - \frac{205}{96} \pi^2 \right] \nu - \frac{155}{96} \nu^2 - \frac{35}{5184} \nu^3 \right) x^3 \right. \nonumber \\ &\qquad+ \left. \left(-\frac{3969}{128} + \nu \,e_4(\nu) + \frac{448}{15}\nu\ln x\right) x^4\right. \nonumber \\ &\qquad+ \left. \left(-\frac{45927}{512} + \nu \,e_5(\nu) + \left[-\frac{4988}{35}-\frac{1904}{15}\nu\right]\nu\ln x\right) x^5\right\} \,.
\end{align}
Here $e_4(\nu)$ and $e_5(\nu)$ denote some unknown 4PN and 5PN coefficients which are some polynomials in the symmetric mass ratio $\nu$ --- this can be proved from the fact that the energy function for general orbits (i.e. before restriction to circular orbits) must be a polynomial in the two separate masses $m_1$ and $m_2$. This 5PN accurate formula could be used to compute the location of the innermost circular orbit (ICO) in the comparable mass regime, which also coincides with the innermost stable circular orbit (ISCO) in the extreme mass ratio regime. The shift of the Schwarzschild ISCO due to the conservative part of the self-force has been recently computed \cite{BaSa.09}. A high-order PN comparison with this result would be interesting, but requires at least the evaluation of the coefficients $e_4(\nu)$ and $e_5(\nu)$ in the extreme mass ratio regime, i.e. the knowledge of $e_4(0)$ and $e_5(0)$. 

\section{The gauge invariant redshift observable}\label{secV}

We are now ready to implement our computation of the gauge invariant redshift observable \eqref{uT}. We replace the 4PN and 5PN logarithmic terms in the metric coefficients evaluated on the particle \eqref{deltag1} into \eqref{uT}. We are careful at including also the metric up to 1PN order because of the coupling between 1PN and 4PN orders which produce 5PN terms. The result is valid for any orbit in a general frame. Next we go to the frame of the center-of-mass defined by $G^i=0$, where $G^i$ is the conserved integral of the center of mass. We have found the 4PN and 5PN logarithms in $G^i$ in Eq.~\eqref{deltaGi}, and from this we compute the displacement of the center of mass for circular orbits. As already used in Sec.~\ref{secIV} we find that the first logarithmic terms in the center-of-mass integral for circular orbits arise only at 5PN order. We obtain
\begin{equation}\label{deltaGic}
    \delta G^i = -\frac{324}{7}\,m\,\nu^2\,\Delta\,\gamma^5\,\ln\gamma\;y_{12}^i\,,
\end{equation}
where $\Delta=(m_2-m_1)/m=\sqrt{1-4\nu}$ is the relative mass difference. The correction to the individual center-of-mass positions will thus be given by $\delta y_a^i=-\delta G^i/m$ for $a$=1,2 (see e.g. the Appendix B in Paper I), and similarly $\delta v_a^i = - \delta \dot{G}^i / m$ for the individual center-of-mass velocities. We already notice that because of the factor $\nu^2$ in \eqref{deltaGic} this correction will not influence the SF limit. Next we reduce the latter expression to circular orbits, replacing all orbital frequencies by their expressions in terms of $\gamma$, and then replacing all $\gamma$'s by their expressions in terms of $x$. The formulas \eqref{deltaOm2} and \eqref{deltagam} for the 4PN and 5PN logarithms play of course the crucial role. Finally we end up with the full correction due to the 4PN and 5PN logarithmic terms for circular orbits in our redshift observable $u^T$ as (removing now the index 1 indicating the smaller mass) 
\begin{equation}\label{uTpn}
    \delta u^T = \left[-\frac{32}{5} - \frac{32}{5}\Delta +\frac{64}{15}\nu +  \left(\frac{478}{105}+\frac{478}{105}\Delta+\frac{1684}{21}\nu+\frac{4388}{105}\Delta\nu-\frac{3664}{105}\nu^2\right)x\right] \nu\,x^5\,\ln x\,.
\end{equation}
This correction is valid for any mass ratio $q=m_1/m_2$ and is to be added to the 3PN expression for $u^T$ obtained in Eq.~(4.10) of Paper I. Being proportional to the symmetric mass ratio $\nu$, the correction \eqref{uTpn} vanishes in the test-mass limit, which is to be expected since the Schwarzschild result for $u^T(\Omega)$ does not involve any logarithm.

We now investigate the small mass ratio regime $q\ll 1$. As in Paper I we introduce a convenient PN parameter appropriate to the small mass limit of particle 1:
\begin{equation}\label{y}
    y \equiv \left( \frac{G\,m_2\,\Omega}{c^3} \right)^{2/3}\,.
\end{equation}
We immediately obtain, up to say the quadratic order in $q$, and keeping only the relevant logarithmic terms,
\begin{equation}\label{deltauT}
    \delta u^T = q\left[-\frac{64}{5} +  \left(\frac{956}{105}+\frac{4588}{35}q\right)y\right] y^5\,\ln y + \mathcal{O}(q^3)\,.
\end{equation}
Our complete redshift observable, expanded through post-self-force order, is of the type
\begin{equation}\label{utexp}
    u^T = u^T_\mathrm{Schw} + q \, u^T_\mathrm{SF}  + q^2 \, u^T_\mathrm{PSF} + \mathcal{O}(q^3) \,,
\end{equation}
where the Schwarzschild result is known in closed form as $u^T_\mathrm{Schw} = \left( 1 - 3 y \right)^{-1/2}$. Adding back the 3PN results of Paper I (see Eq. (5.5) there), we thus find that the self-force contribution is given by\footnote{For clarity we add the Landau $o$ symbol for remainders which takes the standard meaning.}
\begin{align}\label{utSF}
u^T_\mathrm{SF} &= - y - 2 y^2 - 5 y^3 + \left(
- \frac{121}{3} + \frac{41}{32} \pi^2 \right) y^4 \nonumber\\&+ \left(
\alpha_4 - \frac{64}{5}\ln y\right) y^5 + \left(
\alpha_5 + \frac{956}{105}\ln y\right) y^6 + o(y^6)\,.
\end{align}
The expansion \eqref{utSF} was determined up to 2PN order $\propto y^3$ in \cite{De.08} based on the Hadamard-regularized 2PN metric given in \cite{Bl.al.98}. The result at 3PN order $\propto y^4$ was obtained in Paper I using the powerful dimensional regularization (as opposed to Hadamard's regularization which found its limits at that order). By contrast our analytic determination of the logarithmic terms at 4PN and 5PN orders depends only marginally on the regularization scheme.

The coefficients $\alpha_4$ and $\alpha_5$ denote some unknown purely numerical numbers which would be very difficult to compute with PN methods, and should depend crucially on having a consistent regularization scheme. By comparing the expansion \eqref{utSF} with our accurate numerical SF data for $u^T_\text{SF}$, we shall be able to measure these coefficients with at least 8 significant digits for the 4PN coefficient $\alpha_4$, and 5 significant digits for the 5PN coefficient $\alpha_5$. These results, as well as the estimation of even higher-order PN coefficients, will be detailed in Sec.~\ref{secVI}.

Similarly, adding up the results of Paper I for the post-self-force term, we get 
\begin{align}\label{utpSF}
u^T_\mathrm{PSF} &= y + 3 y^2 + \frac{97}{8} y^3 + \left(
        \frac{725}{12} - \frac{41}{64} \pi^2 \right) y^4 \nonumber\\&+ \epsilon_4 \,y^5 + \left(\epsilon_5 + \frac{4588}{35}\ln y\right) y^6 + o(y^6)\,.
\end{align}
Note that there is no logarithm at 4PN order in the post-self-force term, as is also seen from Eq.~\eqref{deltauT}; the next 4PN logarithm would arise at cubic order $q^3$, i.e. at the post-post-SF level. The coefficients $\epsilon_4$ and $\epsilon_5$ in \eqref{utpSF} are unknown, and unfortunately they are expected to be extremely difficult to obtain, not only analytically in the standard PN theory, but also numerically as they require a second-order perturbation SF scheme.

\section{Numerical evaluation of post-Newtonian coefficients}\label{secVI}

In the self-force limit, the SF effect $\uT_\SF$ on the redshift observable $\uT$ is related to the regularized metric perturbation $\hR$ at the location of the particle through
\begin{equation}
 u^T_\SF =
    \frac12 (1 - 3y)^{-1/2} \, \ubar^\alpha \ubar^\beta h^\R\lab \, ,
\label{uTeqn}
\end{equation}
where $\ubar^\alpha$ is the background four-velocity of the particle. Beware that here $\hR$ stands in fact for the perturbation \textit{per unit mass ratio}, denoted $\hR/q$ in Paper I (cf. Eq.~(2.11) there). In SF analysis, the combination $\uuh$ arises more naturally than $\uT_\SF$; this is the quantity we shall be interested in fitting in this Section. However our final results in Table \ref{bestfit} will include the corresponding values of the coefficients for the redshift variable $\uT_\SF$. We refer to Sec. II of Paper I for a discussion of the computation of the regularized metric perturbation $\hR$, and the invariant properties of the combination $\uuh$ with respect to the choice of perturbative gauge. In this Section we often use $r = 1/y$, a gauge invariant measure of the orbital radius scaled by the black hole mass $m_2$ [see Eq.~\eqref{y}].

Our earlier numerical work, partially reported in \cite{De.08} and in Paper I, provided values of the function $\uuh(r)$ which cover a range in $r$ from $4$ to $750$. Following a procedure described in \cite{De.al.03}, we have used Monte Carlo analysis to estimate the accuracy of our values for $\uuh$. As was reported in Paper I, this gives us confidence in these base numbers to better than one part in $10^{13}$. We denote a standard error $\sigma$ representing the numerical error in $\uuh$ by
\begin{equation}
  \sigma \simeq  |\uuh| \times {\rm E}\times10^{-13},
\label{asymp}
\end{equation}
where $\rm E \simeq 1$ is being used as a placeholder to identify our estimate of the errors in our numerical results.

\subsection{Overview}
\label{secVIA}

A common task in physics is creating a functional model for a set of data. In our problem we have a set of $N$ data points $f_i$ and associated uncertainties $\sigma_i$, with each pair evaluated at an abscissa $r_i$. We wish to represent this data as some model function $f(r)$ which consists of a linear sum of $M$ basis functions $F_j(r)$ such that
\begin{equation}
  f(r) = \sum_{j=1}^M c_j F_j(r) \, .
\end{equation}
The numerical goal is to determine the $M$ coefficients $c_j$ which yield the best fit in a least squares sense over the range of data. That is, the $c_j$ are to be chosen such that
\begin{equation}
  \chsq \equiv \sum_{i=1}^N
       \left[\frac{f_i - \sum_{j=1}^M c_j F_j(r_i)}{\sigma_i}\right]^2
\label{chi2def}
\end{equation}
is a minimum under small changes in the $c_j$. For our application we choose the basis functions $F_j(r)$ to be a set of terms which are typical in PN expansions, such as $r^{-1}$, $r^{-2}$, \ldots, and also terms such as $r^{-5}\ln(r)$.

Our analysis depends heavily upon Ref.~\cite{PrTeVeFl}; we use both the methods and the computer code for solving systems of linear algebraic equations as described therein.  While we do employ standard, least squares methods for solving a system of linear equations to determine the $c_j$, we also recognize that a solution to this extremum problem is not guaranteed to provide an accurate representation of the data $(r_i, f_i,\sigma_i)$. The quality of the numerical fit is measured by $\chsq$ as defined in Eq.~(\ref{chi2def}). If the model of the data is a good one, then the $\chi^2$ statistic itself has an expectation value of the number of degrees of freedom in the problem, $N-M$, with an uncertainty (standard deviation) of $\sqrt{2(N-M)}$. In particular, a large residual $\chsq$ would correspond to under-fitting the data whereas a $\chsq$ that is too small corresponds to over-fitting the data, which amounts to fitting randomness in the residuals.

The numerical evaluation of the fitting coefficient $c_j$ includes a determination of its uncertainty $\Sigma_j$ which depends upon i) the actual values of $r_i$ in use, ii) all of the $\sigma_i$, and iii) the set of basis functions $F_j(r)$. In fact, the estimate of the $\Sigma_j$ depends solely upon the design matrix
\begin{equation}
  A_{ij} \equiv \frac{F_j(r_i)}{\sigma_i} \, ,
\end{equation}
and not at all on the data (or residuals) being fitted. However, the estimates of the $\Sigma_j$ are only valid if the data are well represented by the set of basis functions. For emphasis: the $\Sigma_j$ depend upon $F_j(r_i)$ and upon $\sigma_i$ but \textit{are completely independent} of the $f_i$. Only if the fit is considered to be good, could the $\Sigma_j$ give any kind of realistic estimate for the uncertainty in the coefficients $c_j$. If the fit is not of high quality (unacceptable $\chsq$), then the $\Sigma_j$ bear no useful information \cite{PrTeVeFl}. We will come back to this point in the discussion below.

A further remark concerning the meaning of the $\Sigma_j$ is appropriate. Fitting the data as described to determine the coefficients is a standard, well defined statistical procedure. If we were to change the integration routine used to generate the $\uuh(r_i)$, which are the set of input data values $f_i$, with the restriction that we maintained the same numerical accuracy then the $f_i$ would each change in a random way governed by $\sigma_i$. If the coefficients $c_j$ were then determined for this second data set, the statistical analysis ensures that the $\Sigma_j$ associated with this second data set would be identical to those of the first set and the newly determined estimates of the $c_j$ would differ from the initial ones in a statistical fashion governed by the $\Sigma_j$. Recall that the $\Sigma_j$ depend upon the choice of the $r_i$, upon the $\sigma_i$ for the individual data points and upon the set of basis functions $F_j(r)$. The $\Sigma_j$ are completely independent of the data values $f_i$.

Now we consider two other possible changes. If we add an extra data point, or if we add another basis function not orthogonal to the others (this would be typical over a finite set of data points, unless we carefully engineered otherwise) the design matrix changes accordingly, all estimated coefficients $c_j$ change accordingly, and the estimated $\Sigma_j$ change in ways which are not easily related to the previous results. In particular, if we add an additional basis function $F_{M+1}$ to the previous set, so there is now one more coefficient $c_{M+1}$ to be fit, and we compare the first $M$ values of the new $c_j$ to their earlier values, their differences need not be closely related to either the first or second set of $\Sigma_j$. Thus, a change in the design matrix of the problem leads to an inability to make any intuitive prediction about what to expect for the new $c_j$, and there is no reason to expect that the differences of the $c_j$ respect the values of the $\Sigma_j$ for these two different statistical problems.

We also should remark that the task of determining coefficients in the $1/r$ characterization of our numerical data is almost incompatible with the task of determining an asymptotic expansion of $\uuh$ from an analytic analysis. Analytically, the strict $r\rightarrow+\infty$ limit is always technically possible, whereas numerically, not only is that limit {\it never} attainable, but we must always contend with function evaluations at just a finite number of discrete points, obtained within a finite range of the independent variable, and computed with finite numerical precision.  Nevertheless, this is what we intend to do.

In practice, the numerical problem is even more constrained. At large $r$, even though the data may still be computable there, the higher order terms for which we are interested in evaluating PN coefficients rapidly descend below the error level of our numerical data. This is clearly evident in Fig.~\ref{bestfig} below. For small $r$, the introduction of so many PN coefficients is necessary that it becomes extremely difficult to characterize our numerical data accurately. Thus, in practice, we find ourselves actually working with less than the full range of our available data. At large $r$ we could effectively drop points because they contribute so little to any fit we consider. At the other extreme, the advantage of adding more points in going to smaller $r$ is rapidly outweighed by the increased uncertainty in every fitted coefficient. This results from the need to add more basis functions in an attempt to fit the data at small $r$. Further details will become evident in Sec.~\ref{secVID} below.

\subsection{Framework for evaluating PN coefficients numerically}
\label{secVIB}

In a generic fashion we describe an expansion of $\ubar^\alpha \ubar^\beta h^\R\lab$ in terms of PN coefficients $a_j$ and $b_j$ with
\begin{equation}
	\uuh = \sum_{j\geqslant0} \frac{a_j}{r^{j+1}} - \ln r \sum_{j\geqslant4} \frac{b_j}{r^{j+1}} \, ,
\end{equation}
where $a_0$ is the Newtonian term, $a_1$ is the 1PN term and so on. Similarly, for use in applications involving $u^T$ we also introduce the coefficients $\alpha_j$ and $\beta_j$ in the expansion of the SF contribution
\begin{equation}
	u^T_\text{SF} = \sum_{j\geqslant0} \frac{\alpha_j}{r^{j+1}}  - \ln r \sum_{j\geqslant4} \frac{\beta_j}{r^{j+1}} \, .
\end{equation}
These series allow for the possibility of logarithmic terms, which are known not to start before the 4PN order. We also concluded in Sec.~\ref{secII} that $(\ln{r})^2$ terms cannot arise before the 5.5PN order. Since we are computing a conservative effect, possible time-odd logarithmic squared contributions at the 5.5PN or 6.5PN orders do not contribute. But there is still the possibility for a conservative 7PN $(\ln{r})^2$ effect, probably originating from a tail modification of the dissipative 5.5PN $(\ln{r})^2$ term. However, we shall not permit for such a small effect in our fits. As discussed below in Sec.~\ref{secVID}, we already have problems distinguishing the 7PN linear $\ln{r}$ term from the 7PN non-logarithmic contribution.

The analytically determined values of the coefficients $a_0$, $a_1$, $a_2$, $a_3$ and $\alpha_0$, $\alpha_1$, $\alpha_2$, $\alpha_3$ computed in Ref.~\cite{De.08} and Paper I are reported in Table \ref{known}, together with the new results $b_{4}=-\frac{128}{5}$, $b_{5}=\frac{5944}{105}$ and $\beta_{4}=-\frac{64}{5}$, $\beta_{5}=\frac{956}{105}$ of the present work. 

\begin{table*}[h]
	\begin{center}
		\begin{tabular}{c | c || c | c}
			\hline\hline
			coeff. & value & coeff. & value \\
			\hline
			$a_0$ & $-2$ & $\alpha_0$ & $-1$ \\
			$a_1$ & $-1$ & $\alpha_1$ & $-2$ \\
			$a_2$ & $-\frac{7}{4}$ & $\alpha_2$ & $-5$ \\
			$a_3$ & $-\frac{1387}{24} + \frac{41}{16}\pi^2$ & $\alpha_3$ & $-\frac{121}{3}+\frac{41}{32}\pi^2$ \\
			$b_4$ & $-\frac{128}{5}$ & $\beta_4$ & $-\frac{64}{5}$ \\
			$b_5$ & $+\frac{5944}{105}$ & $\beta_5$ & $+\frac{956}{105}$ \\
			\hline\hline
		\end{tabular}
		\caption{The analytically determined PN coefficients for $\uuh$ (left) and $u^T_\text{SF}$ (right).} \label{known}
	\end{center}
\end{table*}

\subsection{Verifying analytically determined PN coefficients}
\label{secVIC}

In this Section we investigate the use of our data for $\uuh$ and the fitting procedures we have described above (and expanded upon in the beginning of Sec.~\ref{secVID}).  We will begin by fitting for enough of the other PN coefficients to be able to verify numerically the various coefficients $a_3$, $b_4$ and $b_5$ now known from PN analysis.  We choose a starting point for the inner boundary of the range, and each range continues out to $r=700$. The results of a series of fits are displayed in Tables \ref{a3fit} and \ref{1stfit}. First we remark that bringing the outer boundary inward as far as to $300$ has very little effect on the outcome of any of these fits, except that the $\chi^2$ statistic decreases as expected with the number of degrees of freedom.

As a first step in this Section, we will complete the task we began in Paper I, namely, the numerical determination of the coefficient $a_3$ (and $\alpha_3$), this time taking fully into account the known logarithmic terms at 4PN and 5PN order. For illustrative purposes only, these results are given in Table \ref{a3fit}. We were able to obtain a fit with six undetermined parameters, and could include data from $r=700$ down to $r=35$.  Note that, with the inclusion of the $b_4$ and $b_5$ coefficients, the precision of our tabulated value for $a_3$ has increased by more than four orders of magnitude from Paper I, although our accuracy is still no better than about $2\Sigma$. Such a discrepancy is not uncommon. The uncertainty, $\Sigma$, reflects only how well the data in the given, finite range can be represented by a combination of the basis functions. It is not a measure of the quality of a coefficient when considered as a PN expansion parameter, which necessarily involves an $r\rightarrow+\infty$ limiting process.

\begin{table*}[h]
\begin{center}
\begin{tabular}{c | l l}
\hline\hline
coeff. &~~& \hspace{0.7cm} value \\
\hline
$a_3$  && $-32.5008069(7)$ \\
$a_4$  && $-121.30254(30)$ \\
$a_5$  && $-42.99(5)$ \\
$a_6$  && $-228(6)$   \\
$b_6$  && $+677(2)$   \\
$a_7$  && $-8226(27)$ \\
\hline\hline
\end{tabular}
\caption{
The results of a numerical fit for a set of coefficients which includes the analytically known $a_3$. Thus this is \textit{not} the best-fit of our data possible, but it allows for a comparison with Table \ref{bestfit}. The uncertainty in the last digit or two is in parentheses. The range runs from $r=35$ to $r=700$, with 266 data points and a respectable $\chsq$ of 264.}
\label{a3fit}
\end{center}
\end{table*}

Our next step is to include the known value for $a_3$ and to use our numerical data to estimate values for the $b_4$ and $b_5$ coefficients. Our best quality numerical result was obtained with five fitted parameters, over a range from $r=700$ down to only $r=65$, and is given in the first row of Table \ref{1stfit}.  Notice that while our $b_4$ is determined relatively precisely, it has only about $6\Sigma$ accuracy.  The higher order coefficient $b_5$ is more difficult to obtain and, at this point, it is very poorly determined. It corresponds to a term which falls off rapidly with increasing $r$ and is significant over a relatively small inner part of the fitted range.  

We can of course use the known value of $b_4$ in order to improve the accuracy for $b_5$.  If we do this without adding another parameter to fit, we immediately get a fit of very poor quality, since we have moved $b_4$ far from its best-fit value; as shown in the second row of Table \ref{1stfit}, we must move the inner boundary out to $r=85$ to re-establish a good fit.

\begin{table*}[h]
	\begin{center}
		\begin{tabular}{c | l | l | l | l | l | l | l | l | l}
			\hline\hline
			$r_\text{min}$ & deg. & \hspace{0.04cm} $\chi^2$ & \hspace{0.8cm} $a_4$ & \hspace{0.9cm} $b_4$ & \hspace{0.6cm} $a_5$ & \hspace{0.5cm} $b_5$ & \hspace{0.6cm} $a_6$ & \hspace{0.3cm} $b_6$ & \hspace{0.7cm} $a_7$ \\
			\hline
  			65 & 231 & 222 & $-121.40(1)$ & $-25.6116(20)$ & $-102(1)$ & $45.5(3)$ & $-2081(9)$ &&\\ 
			85 & 212 & 207 & $-121.3180(7)$ & & $-91.45(70)$ & $48.48(15)$ & $-2170(8)$ &&\\ 
			65 & 231 & 222 & $-121.313(1)$  & & $-79(2)$ & $50.6(4)$ & $-1868(44)$ & $131(21)$ \\ 
			40 & 255 & 247 & $-121.3052(6)$ & & $-47(1)$ & $55.7(2)$ & $-359(41)$ & $625(15)$ & $-7722(162)$ \\ 
			\hline\hline
		\end{tabular}
		\caption{The numerically determined PN coefficients for $\uuh$. Each row represents a different fit. The first three columns give the starting point $r_\text{min}$ at the inner boundary of the fitting range, the degrees of freedom of the fit, $N-M$, and the $\chi^2$ statistic for the chosen fit. If a value for a coefficient is not shown, then either that parameter was not included in that particular fit (far right) or its analytically known value was used (e.g., $b_4$). The formal uncertainty of a coefficient in the last digit or two is in parentheses. The outer boundary is $700$ in each case.} \label{1stfit}
	\end{center}
\end{table*}

The inclusion of basis functions for the higher order coefficients, $b_6$ and $a_7$, as shown in the third and fourth rows, respectively, allows the inner boundary for the fit to move to smaller $r$ where the higher PN terms are more important. The third row of the table shows that adding another parameter allows us to move the inner boundary to $r=65$, while the final row shows that we can now add one further fitted parameter, and obtain a good quality fit by pushing the inner boundary to $r=40$. Only in this row is the $b_5$ parameter close to its known value, but it is still off by around $4.5\Sigma$ (see Table~\ref{analytic} below). Moreover, we have reached a limit for treating our data in this way, adding further parameters and inner points does not result in any higher quality fit.

By now we have presented enough to show that we have data which allows high precision, with an accuracy that we now have some experience in relating to the computed error estimates. This experience will be valuable when we come to discuss further results in the next Section. For convenience, we summarize the relevant information further, in Table \ref{analytic}, referring just to our estimates of known PN parameters, and relating our error estimates to the observed accuracy.

\begin{table*}[h]
\begin{center}
\begin{tabular}{l | c | l | l | l }
\hline\hline
source & coeff. & \hspace{0.5cm} estimate  & accuracy & \hspace{0.3cm} exact result\\
\hline
Paper I           & $\alpha_3$   & $-27.677(5)$     & $\rightarrow(11) $ &
$-27.6879\cdots$    \\
Table \ref{a3fit} & $a_3$        & $-32.5008069(7)$ & $\rightarrow(15) $ &
$-32.5008054\cdots$ \\
Table \ref{1stfit} & $b_4$        & $-25.6116(20)$    & $\rightarrow(116)$ &
$-25.6$             \\
Table \ref{1stfit} & $b_5$        & $+55.7(2)$        & $\rightarrow(9)  $ &
$+56.6095\cdots$    \\
\hline\hline
\end{tabular}
\caption{Comparing the analytically known PN coefficients (column 5) with their numerically determined counterparts (column 3), and comparing the numerically determined error estimates (column 3) with the apparent accuracy (column 4). The source of the data is given in column 1.} \label{analytic}
\end{center}
\end{table*}

\subsection{Determining higher order PN terms numerically}\label{secVID}

In this Section we turn our attention to using our numerical SF data and fitting procedures to obtain as many as possible unknown PN coefficients, by making maximum use of the coefficients which are already known. We find that in our \textit{best fit} analysis we can use a set of five basis functions corresponding to the unknown coefficients $a_4$, $a_5$, $a_6$, $b_6$ and $a_7$.

\begin{figure}
    \includegraphics[clip,angle=0,width=15.5cm,]{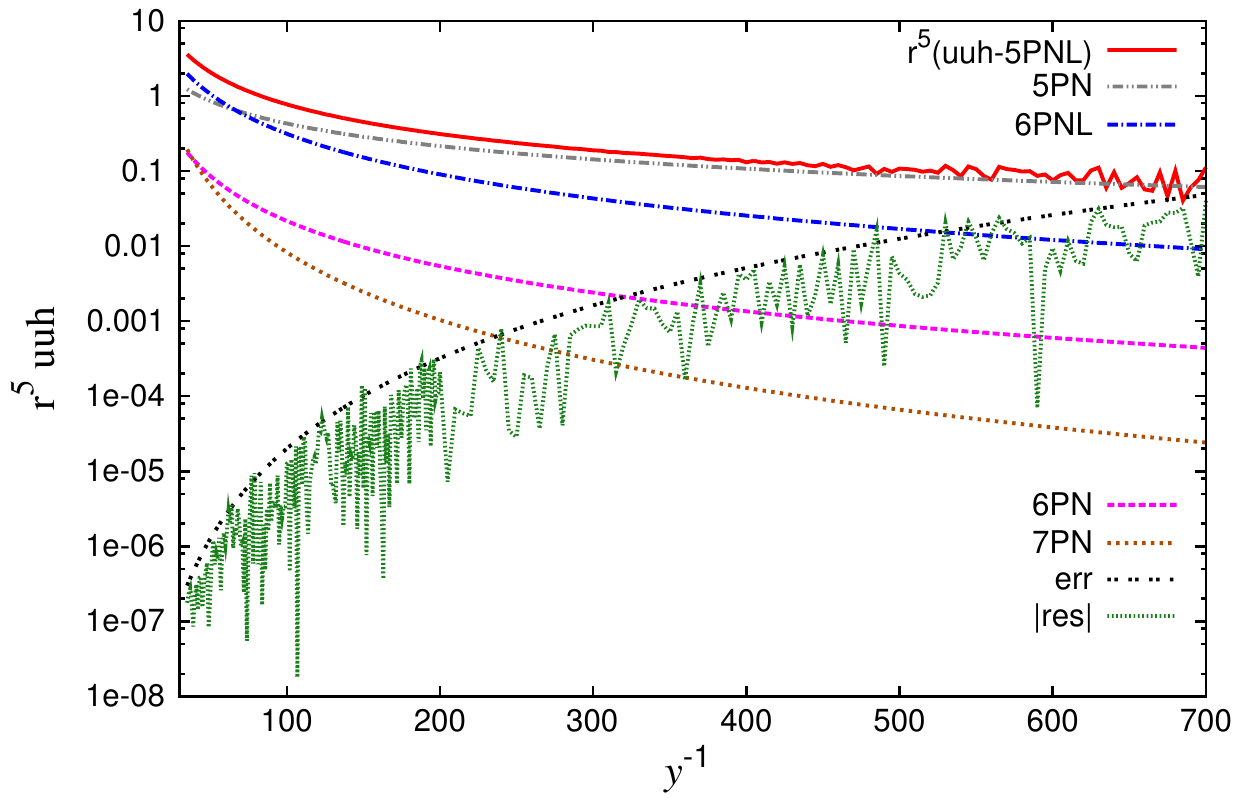}
\caption{The absolute value of the contributions of the numerically determined post-Newtonian terms to $r^5\uuh$. Here PNL refers to just the logarithm term at the specified order. The contribution of $a_4$ is not shown but would be a horizontal line (since the 4PN terms behaves like $r^{-5}$) at approximately 121.3\,. The remainder after $a_4$ and all the known coefficients are removed from $r^5\uuh$ is the top (red) continuous line. The lower (black) dotted line labelled ``err'' shows the uncertainty in $r^5\uuh$, namely $2{\rm E} \, r^4 \times10^{-13}$. The jagged (green) line labelled ``$|$res$|$'' is the absolute remainder after all of the fitted terms have been removed. The figure reveals that, with regard to the uncertainty of the calculated $\uuh$, the choice $E\simeq 1$ was slightly too large.}
\label{bestfig}
\end{figure}

\begin{table*}[h]
	\begin{center}
		\begin{tabular}{c | l || c | l}
			\hline\hline
			coeff. & \hspace{0.5cm} value & coeff. & \hspace{0.5cm} value \\
			\hline
			$a_4$ & $-121.30310(10)$ & $\alpha_4$ & $-114.34747(5)$ \\
			$a_5$ & $-42.89(2)$      & $\alpha_5$ & $-245.53(1)$ \\                 
			$a_6$ & $-215(4)$        & $\alpha_6$ & $-695(2)$ \\                 
			$b_6$ & $+680(1)$        & $\beta_6$  & $+339.3(5)$ \\
			$a_7$ & $-8279(25)$      & $\alpha_7$ & $-5837(16)$ \\                 
			\hline\hline
		\end{tabular}
		\caption{The numerically determined values of higher-order PN coefficients for $\uuh$ (left) and for $u_\text{SF}^T$ (right). The uncertainty in the last digit or two is in parentheses. The range runs from $r=40$ to $r=700$, with 261 data points being fit. The $\chsq$ statistic is 259. We believe that a contribution from a $b_7$ term piggybacks on the $a_7$ coefficient. Both terms fall off rapidly and have influence over the fit only at small $r$. And the radial dependence of these two terms only differ by a factor of $\ln r$ [or possibly $(\ln r)^2$] which changes slowly over their limited range of significance.} \label{bestfit}
	\end{center}
\end{table*}
In Table \ref{bestfit}, we describe the numerical fit of our data over a range in $r$ from $40$ to $700$. The $\chsq$ statistic is 259 and slightly larger than the degrees of freedom, 256, which denotes a good fit. Further, we expect that a good fit would be insensitive to changes in the boundaries of the range of data being fit, and we find, indeed, that if the outer boundary of the range decreases to $300$ then essentially none of the data in the Table changes, except for $\chsq$ and the degrees of freedom which decrease in a consistent fashion. Figure \ref{bestfig} shows that in the outer part of the range $\uuh$ is heavily dominated by only a few lower order terms in the PN expansion --- those above the lower black double-dashed line in the figure.

The inner edge of the range is more troublesome. The importance of a given higher order PN term decreases rapidly with increasing $r$. Moving the inner boundary of the range outward might move a currently well determined term into insignificance. This could actually lead to a smaller $\chsq$, but it would also lead to an increase in the $\Sigma_j$ of every coefficient. Moving the inner edge of the range inward might require that an additional higher order term be added to the fit. This extra term loses significance quickly with increasing $r$ so the new coefficient will be poorly determined and also result in an overall looser fit with an increase of $\Sigma_j$ for all of the coefficients. If the inner boundary and the set of basis functions are chosen properly, then a robust fit is revealed when the parameters being fit are insensitive to modest changes in the boundaries of the range. The fit described in Table \ref{bestfit} appears to be robust. The parameters in this Table are consistent with all fits with the inner boundary of the range varying from 35 to 45 and the outer boundary varying from 300 to 700.

If an additional term, with coefficient $b_7$, is added to the basis functions then, for identical ranges, each of the $\Sigma_j$ increases by a factor of about ten, and the changes in $a_4$ and $a_5$ are within this uncertainty. The coefficient $a_6$ changes sign and $b_6$ and $a_7$ change by an amount significantly larger than the corresponding $\Sigma_j$. And the new coefficient $b_7$ is quite large. In the context of fitting data to a set of basis functions these are recognized symptoms of over-fitting and imply that the extra coefficient degrades the fit.

How should we (and others) interpret the data in Table \ref{bestfit}? To guide our discussion of this very important question, we assemble together into Table \ref{numerical} all the relevant results from the earlier fits of Sec.~\ref{secVIC} which relate to the best prior estimates we have there for $a_4$, $a_5$, $a_6$, $b_6$ and $a_7$ which we have finally calculated here. As was shown in Table \ref{analytic} and is now evident in Table \ref{numerical}, our numerical accuracy tends to be in the range of $2-6 \Sigma$, both when comparing the best results for $a_4$, $a_5$, $a_6$, $b_6$ and $a_7$ from Sec.~\ref{secVIC} with those obtained here and, we would suggest, for the purposes of comparing the results of this Section with future PN coefficients.

\begin{table*}[h]
\begin{center}
\begin{tabular}{c | l | l | l }
\hline\hline
coeff. & \hspace{0.00cm} Table \ref{bestfit} (best) & \hspace{1.1cm} Table \ref{a3fit} & \hspace{0.9cm} Table \ref{1stfit}  \\
\hline
 $a_4$   & $-121.30310(10)  $   & $-121.30254(30) \rightarrow(56) $   &
$-121.3052(6)\rightarrow(21) $  \\
 $a_5$   & $-42.89(2)       $   & $-42.99(5)      \rightarrow(10) $   & $-47(1)
     \rightarrow(4) $  \\
 $a_6$   & $-215(4)          $   & $-228(6)       \rightarrow(13) $   &
$-359(41)    \rightarrow(144) $  \\
 $b_6$   & $+680(1)          $   & $+677(2)       \rightarrow(3) $   &
$+625(15)    \rightarrow(55) $  \\
 $a_7$   & $-8279(25)      $   & $-8226(27)       \rightarrow(53)$   &
$-7722(162)  \rightarrow(557) $  \\
\hline\hline
\end{tabular}
\caption{Comparing the ``best fit'' numerical values and statistical uncertainties of the estimated PN coefficients in Table \ref{bestfit} to other numerical evaluations of these same quantities in Sec.~\ref{secVIC}.}
\label{numerical}
\end{center}
\end{table*}

\subsection{Summary}\label{secVIE}

\begin{figure}
    \includegraphics[width=10.5cm,angle=-90]{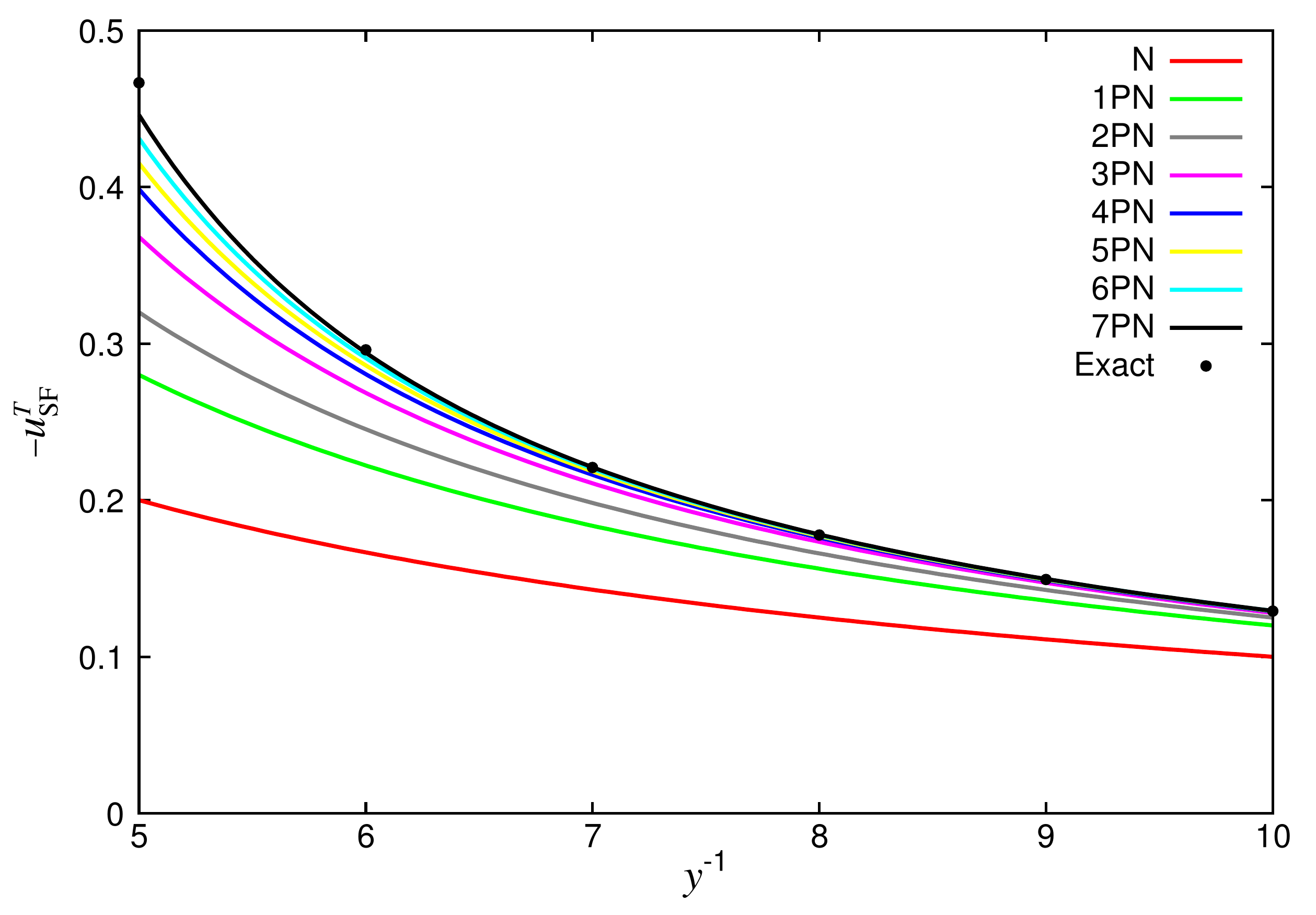}
    \caption{\footnotesize The self-force contribution $u^T_\mathrm{SF}$ to $u^T$ plotted as a function of the gauge invariant variable $y^{-1}$. Note that $y^{-1}$ is an invariant measure of the orbital radius scaled by the black hole mass $m_2$ [see Eq. \eqref{y}]. The ``exact'' numerical points are taken from Ref.~\cite{De.08}. Here, PN refers to all terms, including logarithms, up to the specified order (however recall that we did not include in our fit a log-term at 7PN order).}
    \label{uT_SF}
\end{figure}
Our best fit can be visualized in Fig.~\ref{uT_SF}, where we plot the self-force effect $u_\text{SF}^T$ on the redshift variable $u^T$ as a function of $r=y^{-1}$, as well as several truncated PN series up to 7PN order, based on the analytically determined coefficients summarized in Table \ref{known}, as well as our best fit of the higher-order PN coefficients reported in Table \ref{bestfit}. Observe in particular the smooth convergence of the successive PN approximations towards the exact SF results. Note, though, that there is still a small separation between the 7PN curve and the exact data in the very relativistic regime shown at the extreme left of Fig.~\ref{uT_SF}.

We have found that our data in the limited range of $35 \leqslant r \leqslant 700$ can be extremely well characterized by a fit with five appropriately chosen (basis) functions. That is, the coefficients in Table \ref{bestfit} are well determined, with small uncertainties, and small changes in the actual details of the fit result in coefficients lying within their error estimates. Fewer coefficients would result in a very poor characterization of the same data while more coefficients result in large uncertainties in the estimated coefficients, which themselves become overly sensitive to small changes in specific details (such as the actual choice of points to be fitted). In practice, over the data range we finally choose, and with the five coefficients we fit for, we end up with exceedingly good results for the estimated coefficients, and with residuals which sink to the level of our noise. We have a very high quality fit which is quite insensitive to minor details. Nevertheless, as Tables \ref{analytic} and \ref{numerical} hint, error estimates for these highest order coefficients should be regarded with an appropriate degree of caution.

\section*{Acknowledgements}

SD and BFW acknowledge support through grants PHY-0555484 and PHY-0855503 from the National Science Foundation. LB and ALT acknowledge support from the Programme International de Coop\'eration Scientifique (CNRS--PICS). 

\appendix

\section{Formulas to compute the PN logarithms}\label{appA}
 
In Sec.~\ref{secIII} we looked for poles generating near-zone logarithms when integrating the field equations at quadratic non-linear order. We used the propagator of the ``instantaneous'' potentials defined by
\begin{equation}\label{propa}
\mathcal{I}^{-1} \equiv \mathop{\mathrm{FP}}_{B=0}\,\sum_{k=0}^{+\infty}\left(\frac{\partial}{c\partial t}\right)^{2k}\Delta^{-k-1}\left(\frac{r}{\lambda}\right)^B\,,
\end{equation}
and acting on a source term of the type $r^{-2}F(\mathbf{n},u)$ where $u=t-r/c$; see Eq.~\eqref{deltah2}. We consider here a single multipolar piece in the source term, say $r^{-2}\hat{n}_L F(u)$. The function $F$ is typically a product of the mass with some time derivatives of multipole moments. We recall that the propagator \eqref{propa} depends on the length scale $\lambda=cP$, where $P$ is the period of the source; we thus consider
\begin{equation}\label{sol}
\Phi_L = \mathcal{I}^{-1}\left[\frac{\hat{n}_L}{r^2}F(u)\right] \,.
\end{equation}
In this Appendix we shall provide a general and compact formula giving all the logarithms in the near-zone expansion of the solution \eqref{sol}. The logarithms come from expanding the retardation $u=t-r/c$ in the source when $r/c\rightarrow 0$, integrating each of the terms using the formulas \eqref{mathieu}--\eqref{alphaB}, and finally taking the finite part (FP) associated with the poles $\propto B^{-1}$. Our compact formula gives the result of all these operations as
\begin{equation}\label{log}
\delta\Phi_L = \frac{(-c)^{\ell+1}}{2}\ln\left(\frac{r}{\lambda}\right)\hat{\partial}_L\left\{\frac{F^{(-\ell-1)}(t-r/c)-F^{(-\ell-1)}(t+r/c)}{r}\right\}\,,
\end{equation}
where $F^{(-\ell-1)}$ denotes the $(\ell+1)$-th time anti-derivative of the function $F$. By $\delta\Phi_L$ we mean the contribution of logarithms in $\Phi_L$; thus all the other terms in $\Phi_L$ besides $\delta\Phi_L$ admit an expansion when $r\rightarrow 0$ in simple powers of $r$ without logarithms. Note that the factor of the logarithm in Eq.~\eqref{log} is a multipolar antisymmetric homogeneous solution of the wave equation which is regular at the origin, when $r\rightarrow 0$. The logarithms in \eqref{log} are thus of the NZ type; no FZ logarithms are generated from a source term $r^{-2}\hat{n}_L F(u)$. We recall also from Sec.~\ref{secII} that the FZ logarithms start to arise at the cubic $n=3$ non linear iteration, and that they do not contribute to the conservative part of the dynamics of compact binaries.

The formal near-zone expansion of $\delta\Phi_L$ reads
\begin{equation}\label{logexp}
\delta\Phi_L = (-)^\ell\ln\left(\frac{r}{\lambda}\right)\sum_{i=0}^{+\infty}\frac{\hat{n}_L \,r^{2i+\ell}}{2^ii!(2i+2\ell+1)!!}\frac{F^{(2i+\ell)}(t)}{c^{2i+\ell}}\,.
\end{equation}
At the 1PN relative order required for our computation in \eqref{deltah2}--\eqref{deltah2exp}, we have
\begin{equation}\label{logexp1PN}
\delta\Phi_L = \frac{(-)^\ell\hat{x}_L}{(2\ell+1)!! c^\ell}\left[F^{(\ell)}(t)+\frac{r^2}{2c^2(2\ell+3)}F^{(\ell+2)}(t)+\mathcal{O}\left(\frac{1}{c^4}\right)\right]\ln\left(\frac{r}{\lambda}\right)\,.
\end{equation}

The result \eqref{log} can be generalized in the following sense that the same type of result will hold also for non-STF sources. Namely, if we define $\delta\Phi_L$ to be non-STF in $L$, i.e. having $n_L=n_{i_1}\cdots n_{i_\ell}$ in place of the STF product $\hat{n}_L$ in \eqref{sol}, then we can easily prove that the log-terms are given by \eqref{log} with $\partial_L=\partial_{i_1}\cdots \partial_{i_\ell}$ in place of the STF product $\hat{\partial}_L$. Of course all the other terms will be different, but the structure of the log-terms will be the same. Then it is trivial to show that the formula applies as well to a product of Minkowskian outgoing null vectors $k^\alpha=(1,\mathbf{n})$ representing the direction of propagation of gravitational waves, and satisfying $\eta_{\alpha\beta}k^\alpha k^\beta=0$. Considering
\begin{equation}\label{solalpha}
\Phi_{\alpha_1\cdots\alpha_\ell} = \mathcal{I}^{-1}\left[\frac{k_{\alpha_1}\cdots k_{\alpha_\ell}}{r^2}F(u)\right] \,,
\end{equation}
where $k_\alpha=(-1,\mathbf{n})$, we find indeed that the contribution of logarithms in the near-zone expansion of this object is given by
\begin{equation}\label{logalpha}
\delta\Phi_{\alpha_1\cdots\alpha_\ell} = \frac{(-c)^{\ell+1}}{2}\ln\left(\frac{r}{\lambda}\right)\partial_{\alpha_1\cdots\alpha_\ell}\left\{\frac{F^{(-\ell-1)}(t-r/c)-F^{(-\ell-1)}(t+r/c)}{r}\right\}\,.
\end{equation}

We use this result to show that a family of logarithms not considered in Sec. \ref{secIII} is actually pure gauge. We showed there that all the 4PN and 5PN near-zone logarithms come from iterating the leading-order $1/r^2$ part of the quadratic source, namely $Q_{2}\ab = \frac{4 M}{c^4}\,{}^{(2)}z_{1}\ab + \frac{k^\alpha k^\beta}{c^2}\sigma$. However we computed only the first term $\propto {}^{(2)}z_{1}\ab$, which is associated with tails, but we left out the second term $\propto k^\alpha k^\beta$. Now thanks to the structure $\propto k^\alpha k^\beta$ the logarithms appear in the form of a gauge transformation and will never contribute to a gauge invariant result. This was already shown at the level of the dominant 4PN log-term in \cite{BlDa.88}. By expanding $\sigma$ on the basis of STF tensors $\hat{n}_L$ (or rather $\hat{n}_{L-2}$) we need only to prove this for each of the individual multipolar pieces in the source which have the structure
\begin{equation}\label{solalphaL}
\Phi_{L-2}^{\alpha\beta} = \mathcal{I}^{-1}\left[k^\alpha k^\beta\frac{\hat{n}_{L-2}}{r^2}F(u)\right] \,.
\end{equation}
Applying \eqref{logalpha} the logarithms are given by
\begin{equation}\label{logalphaL}
\delta\Phi_{L-2}^{\alpha\beta} = \frac{(-c)^{\ell+1}}{2}\ln\left(\frac{r}{\lambda}\right)\partial^\alpha\partial^\beta\hat{\partial}_{L-2}\left\{\frac{F^{(-\ell-1)}(t-r/c)-F^{(-\ell-1)}(t+r/c)}{r}\right\}\,,
\end{equation}
and can readily be put in the form of a gauge transformation with gauge vector
\begin{equation}
\xi_{L-2}^\alpha = \frac{(-c)^{\ell+1}}{4}\ln\left(\frac{r}{\lambda}\right)\partial^\alpha\hat{\partial}_{L-2}\left\{\frac{F^{(-\ell-1)}(t-r/c)-F^{(-\ell-1)}(t+r/c)}{r}\right\}\,.
\end{equation}
Indeed we have $\delta\Phi_{L-2}^{\alpha\beta}=2\partial^{(\alpha}\xi_{L-2}^{\beta)}-\eta^{\alpha\beta}\partial_\mu\xi_{L-2}^\mu$ modulo some terms which are free of logarithms. Therefore the ``seed'' logarithms generated in this way at quadratic order can be removed by a gauge transformation, and we conclude that the whole family of logarithms coming from the iteration at cubic and higher orders can be removed by a non-linear deformation of the gauge transformation, namely by a coordinate transformation. Thus we do not have to consider these logarithms in our computation of a gauge invariant quantity; only those coming from the first term $\propto {}^{(2)}z_{1}\ab$ in $Q_{2}\ab$ will contribute as computed in Sec.~\ref{secIII}.

\bibliography{}

\end{document}